\documentclass[prb,twocolumn,superscriptaddress,
longbibliography,aps]{revtex4-2}

\usepackage{graphicx}
\usepackage{bm}
\usepackage{color}
\usepackage{epstopdf}
\usepackage{amsmath}
\usepackage{amssymb}
\usepackage{epstopdf}

\usepackage[normalem]{ulem}

\usepackage{xfrac}

\usepackage{multirow}

\usepackage[urlcolor=blue,colorlinks=true,citecolor=blue,linkcolor=blue,pdfstartview={FitH},bookmarks=false]{hyperref}

\graphicspath{{fig/}{./fig/}{.}}

\begin{document}

	\title{Landau levels in Weyl semimetal under uniaxial strain}

	\author{Shivam Yadav}
	\email[e-mail: ]{yshivam91@gmail.com}
	\affiliation{Institute of Nuclear Physics, Polish Academy of Sciences, ul. W. E. Radzikowskiego 152, PL-31342 Krak\'{o}w, Poland}
	
	\author{Andrzej Ptok}
	\email[e-mail: ]{aptok@mmj.pl}
	\affiliation{Institute of Nuclear Physics, Polish Academy of Sciences, ul. W. E. Radzikowskiego 152, PL-31342 Krak\'{o}w, Poland}
	
	\date{\today}

	\begin{abstract}
		The external strain can lead to the similar effect to the external applied magnetic field. 
		Such pseudomagnetic field can be larger than typical magnetic fields, what gives the opportunity to experimentally study the Landau levels.
		In this paper we study the effects of uniaxial strain on the Weyl nodes, using continuum and lattice model. 
		In the continuum model we show that the uniaxial strain leads to magnetic field renormalization, which in practice corresponds to the shift of the Landau levels to higher energy.
		We also investigate type-I and type-II Weyl nodes using lattice model.
		In this case, the magnetic field is introduced by the Pierels substitution, while uniaxial strain by the direction dependence of hopping integrals.
		This allowed us to probe the Landau level and system spectrum which takes form of Hofstadter butterfly.
		We show that the renormalization of magnetic field, similar to this observed in the continuum model, emerges.
	\end{abstract}

	\maketitle

	\section{Introduction}

	In the case of the simplest systems with manifested Dirac cones for example graphene with the honeycomb lattice structure, the external strain can lead to the effects similar to the external applied magnetic field~\cite{guinea.katsnelson.10}.
	The strain acts as a ``chiral'' vector potential that couples to Dirac fermions oppositely in the two valleys K and K'.
	Such a pseudomagnetic field can be larger than $300$~T and has been observed through the spectroscopic measurement of the Landau levels (LLs)~\cite{levy.burke.10}.
	However, the linear dispersion relation in the low-energy electronic excitations around the Fermi level, similar to that observed in the graphene, can be also be found in the Weyl semimetals (WSMs).

	One of the characteristics features of the Dirac spectrum is its LLs structure.
	The dependence of the cyclotron frequency on the magnetic field and the spacing of the LLs distinguishes the WSMs from the other nonrelativistic electron systems. 
	Moreover, the zeroth LL with linear dispersion is observed~\cite{tchoumakov.civelli.16}.
	The impact of the strain can be studied by the precise investigation of the LLs.
	The famous chiral anomaly can be induced in the WSM nanowire under torsion~\cite{pikulin.chen.16}.

Additionally, the time dependent mechanical distortion can not only produce pseudo magnetic field but also a pseudo electric field and can have a full electromagnetic response which can be observed in Hall viscosity ~\cite{shapourian.hughes.15, cortijo.ferreiros.15}, anomalous Hall effect ~\cite{vazifeh.franz.13} and resistance of WSM nano wire under torsion ~\cite{pikulin.chen.16}. Moreover, it can also cause LL collapse ~\cite{lee.park.22, arjona.castro.17}.

	Discovery of the three-dimensional Dirac and WSMs~\cite{yan.felser.17,armitage.mele.18}, like TaAs~\cite{lv.xu.15,lv.weng.15,xu.belopolski.15,yang.liu.15,cichorek.bochenek.22}, Na$_{3}$Bi~\cite{wang.sun.12,liu.zhou.14,zhang.liu.14}, or Cd$_{3}$As$_{2}$~\cite{wang.weng.13,borisenko.gibson.14,neupane.xu.14}, provides the opportunity to study of the impact of the strain on the relativistic electronic excitations in the real materials.
	Additionally, these systems can be excellent platform to the study of the topological properties of WSM under external conditions.
	More realistic DFT calculations indicate that the strain strongly affects the Weyl nodes~\cite{ruan.jian.16,ferreira.manesco.21,wu.li.21,li.zhao.22,wu.ke.23}.
	For example, the total number of the realized Weyl nodes in the system can depend on strain.
	Moreover, in some extreme situations strain can lead to total destruction of Weyl nodes, or destroy the topological phase~\cite{mutch.chen.19}.
	Such modification should be also reflected in the transport properties of the WSMs~\cite{jiang.guo.21}.
	The biaxial strain can lead to increasing of the superconducting temperature transition.
	This was observed experimentally for example in type-II WSM MoTe$_{2}$~\cite{yip.lam.23}.

	\begin{figure}[!t]
		\centering
		\includegraphics[width=\columnwidth]{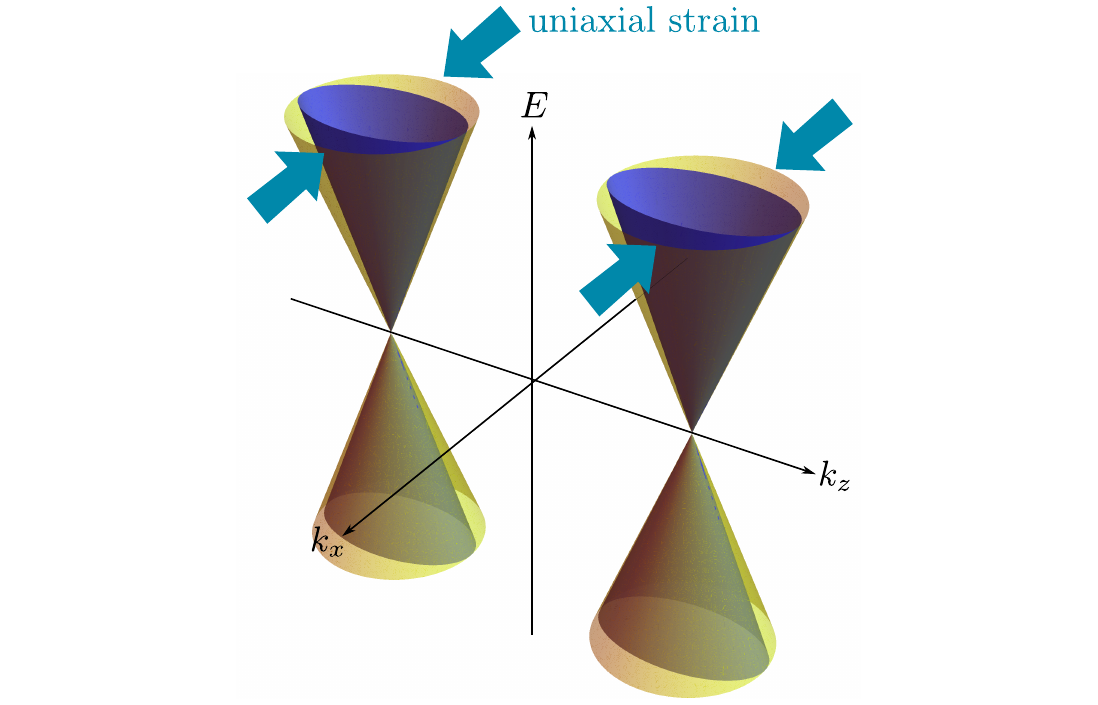}
		\caption{
			Schematic depicts impact of uniaxial strain on dispersion relation of the Weyl semimetal. 
			Yellow cones represent a Weyl node in the absence of external strain, while blue deformed cone represents a Weyl node in the presence of external uniaxial strain (represented by turquoise arrows) along fixed direction. 
			Under uniaxial strain, circular constant energy contour deforms into an ellipse shown by gray solid lines, and elliptical Weyl cone emerge.
			Presented schematic representation correspond directly the system described in Sec.~\ref{sec.cont}.
			\label{fig.schem}
		}
	\end{figure}

	A WSM whe subjected to an inhomogenous strain~\cite{grushin.venderbos.16, pikulin.chen.16} the strain gives rise to a guage which couples with the Weyl fermions which quantizes the spectrum giving rise to Landau level-like structure which are termed as pseudo Landau levels. This gauge can also be introduced by magnetically doping a topological insulator~\cite{liu.ye.13}. Another way to accomplish the chiral pseudo magnetic field is by introduction of screw or edge lattice dislocation of WSM~\cite{sumiyoshi.fujimoto.16}. The inhomogenous strain morphs the separation between the Weyl nodes by making it space dependent. Curl of this space dependence of the separation of Weyl nodes can be interpreted as an axial field analogous to magnetic field. Usually this pseudo magnetic field is incorporated in the effective model of WSM as a perturbation to the linear Hamiltonian~\cite{vazifeh.franz.13, liu.ye.13} and on other occasions by studying the 2D lattice deformation under the application of inhomogenous strain~\cite{shapourian.hughes.15, cortijo.ferreiros.15}.

	In the simplest approximation, we can assume is a homogenous uniaxial strain along some direction (perpendicular to the $z$ axis), leads to the linear deformation of the system within elastic limit.
	Consequently, we can assume that the electron dispersion also exhibits similar deformation (Fig.~\ref{fig.schem}).
	Initially ideal Weyl cone (represented by the yellow surface) is deformed to the elliptical Weyl cone (represented the blue surfaces).
	In our paper, we study the physical properties of the system with pair of elliptical Weyl cones, realized by application of a homogenous uniaxial strain on the system.
	More directly, we discuss the Landau levels spectra within the continuum model (Sec.~\ref{sec.cont}) and system spectra using the lattice model (Sec.~\ref{sec.latt}).
	We show that the uniaxial strain acts on the system as a renormalised magnetic field in both cases.
	Finally, we conclude our findings in Sec.~\ref{sec.sum}.


	\section{Continuum model description}
	\label{sec.cont}

	We start our investigation with the study of the continuum model.
	We begin with the two band low energy effective model for 2 Weyl node in the absence of strain \cite{faruk.23, chan.lee.17}:
	\begin{eqnarray}
		\mathcal{H} &=& t_x \, k_x \sigma_x + t_y \, k_y \sigma_y + w  (k^2_z - \Delta) \sigma_z .
		\label{eq.ham_cont}
	\end{eqnarray} 
	We are working in natural units hence $\hbar = c = 1$. The Weyl nodes appear at $\pm \sqrt{\Delta}$, this makes the separation between the Weyl nodes $2 \sqrt{\Delta}$. It is easily seen that $w$ carries the dimensions of inverse mass along with $t_x$ and $t_y$ being dimensionless.  
	In the elastic limit, the uniaxial strain leads to the linear modification of the displacement between orbitals of neighbouring atoms.
	This feature can be captured by the renormalizing of the hopping integrals ~\cite{cortijo.ferreiros.15}:
	\begin{eqnarray}
		\label{eq.train_k} t_x \rightarrow \gamma_{-} t_x , \quad \text{and} \quad  t_y \rightarrow \gamma_{+} t_y ,
	\end{eqnarray}
	where we introduce $\gamma_{\pm} = 1 \pm \alpha$.
	Here $\alpha$ denotes the parameter controlling strain within $xy$ plane (what is schematically presented in Fig.~\ref{fig.schem}). Additionally, we assume that $t_x = t_y = t$, as we will see later in this  section, this hardly changes the generality of our results.
	Under uniaxial strain, this gauge modifies the Hamiltonian of the system leading to the form:
	\begin{eqnarray}
		\mathcal{H} &=&  t \gamma_- \, k_x \sigma_x + t \gamma_+ \, k_y \sigma_y + w (k^2_z - \Delta) \sigma_z \,.
		\label{eq.ham_str}
	\end{eqnarray}
	As we can see, owing to the introduction of the uniaxial strain $\alpha \neq 0$ the azimuthal dependence of effective Fermi velocity $v ( \varphi ) \sim E ( \textbf{k} ) / k$ becomes obvious, where $\varphi$ denotes angle around the Weyl node.
We should notice, that this clearly shows difference between discussed case of the system under uniaxial strain, where the Fermi velocity is independent by $\varphi$.

	The Weyl nodes are present at $\mathbf{k}_0 = (0, \, 0, \, \chi \sqrt{\Delta})$, where $\chi = \pm 1$. Expanding about the $\mathbf{k}_0$ for small $k_z$ and keeping leading order terms we get,
	\begin{eqnarray}
		\mathcal{H} &=& t \gamma_- \, k_x \sigma_x + t \gamma_+ \, k_y \sigma_y + 2 \chi w \sqrt{\Delta} \, k_z \sigma_z \,,
	\end{eqnarray}
	this shows that $\chi$ manifests as the chirality of the Weyl node separated by $2 \sqrt{\Delta}$ along $k_z$ (as presented schematically on Fig.~\ref{fig.schem}). 
	Now consider an external magnetic field $\textbf{B} = (0, \, 0, \, B)$ and in Landau gauge we can write the vector potential as $\textbf{A} = (0, \, Bx,\,0)$. 
	Using the Pierels's substitution $\mathbf{\Pi} = \mathbf{k} + e \mathbf{A}$, we get the Hamiltonian of the system under uniaxial strain and in presence of magnetic field:
	\begin{eqnarray}
		\mathcal{H} &=& \begin{pmatrix}
			w (\Pi_z^2 - \Delta)  &  t \gamma_- \Pi_x - i t \gamma_+ \Pi_y \\
			t \gamma_- \Pi_x + i t \gamma_+ \Pi_y & - w (\Pi_z^2 - \Delta)
		\end{pmatrix} .
	\end{eqnarray} 
	We wish to find the eigenvalues for this strained Hamiltonian, to accomplish this we assume that the eigenvectors are given by $| \psi_n \rangle = (u_n \; v_n)^{T}$. This gives us the following equations
	\begin{eqnarray}
		\begin{array}{c}
			w (\Pi_z^2 - \Delta) u_n + (t \gamma_- \Pi_x - i t \gamma_+ \Pi_y) v_n = E_n u_n \\ 
			(t \gamma_- \Pi_x + i t \gamma_+ \Pi_y) u_n - w (\Pi_z^2 - \Delta) v_n = E_n v_n
		\end{array} 
		\label{eq:eig_linear}
	\end{eqnarray}
	The equations demand the eigenfunctions to be of form $ | \psi_n \rangle \rightarrow e^{i k_z z} | \psi_n \rangle$, and we define ladder operator $a / l_{B} = t \gamma_- k_x - i t \gamma_+ (k_y + e B x)$. The ladder operators follow the harmonic oscillator algebra given that:
	\begin{eqnarray}
		l_{B} = \frac{1}{ \sqrt{2 t (1-\alpha^2) e B} } .
		\label{eq.mag_len}
	\end{eqnarray}
	As we can see, the magnetic length $l_{B}$ is proportional to $1/\sqrt{1-\alpha^{2}}$.
	For a relatively small value of strain $\alpha$ we can estimate $l_{B}$ as a function proportional to $\alpha^{2}$.

	Using the ladder operators, Eq.~(\ref{eq:eig_linear}) can be rewritten as
	\begin{eqnarray}
			\begin{array}{c}
			a v_n = l_{B} [ E_n - w (k^2_z - \Delta) ] u_n \; , \\ 
			a^{\dagger} u_n = l_{B} [ E_n + w (k^2_z - \Delta) ] v_n \; . 
			\end{array}
	\end{eqnarray}
	Using the equations above we can conclude that 
	\begin{eqnarray}
		\begin{array}{c}
			a a^{\dagger} u_n = { l_B^2 } [ { E_n^2 - w^2 (k^2_z - \Delta)^2 } ] u_n \,, \\
			a^{\dagger} a v_n = { l_B^2 } [ { E_n^2 - w^2 (k^2_z - \Delta)^2 } ] v_n  \,.
		\end{array} 
	\end{eqnarray}
	These eigenfunctions correspond to the harmonic oscillator states and hence the eigen-energies for the WSM under uniaxial strain kept in a magnetic field can be written as:
	\begin{eqnarray}
		E_n^{(\pm)} &=& \pm \sqrt{ \frac{n}{l_B^2} + w^2 (k^2_z - \Delta)^2}
		\label{eq:energy}
	\end{eqnarray}
	here $(\pm)$ represent the conduction and valance bands. We also get
	\begin{eqnarray}
		u_n^{(\pm)} = \sqrt{\frac{1}{2} \left( 1 +  \frac{ w (k^2_z - \Delta) }{ E_n^{(\pm)} } \right)} | n - 1 \rangle ,
	\end{eqnarray}
	and
	\begin{eqnarray}
		v_n^{(\pm)} = \sqrt{\frac{1}{2} \left( 1 -  \frac{ w (k^2_z - \Delta) }{ E_n^{(\pm)} } \right)} |n \rangle .
	\end{eqnarray}

	\begin{figure}[!t]
		\centering
		\includegraphics[width=\linewidth]{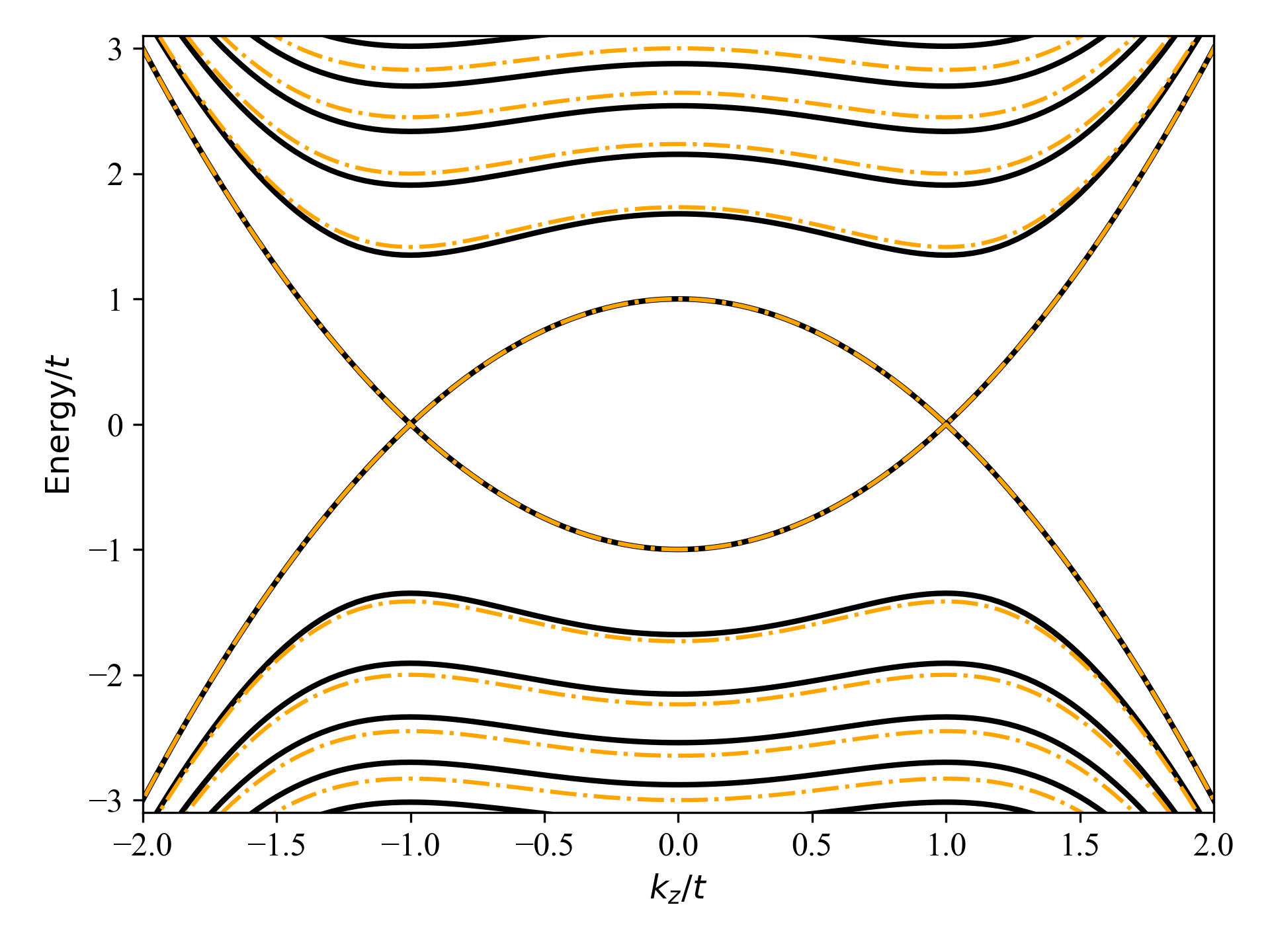}
		\caption{
			Energy dispersion in the ambient and in the presence of strain.
			Orange (dashed) and black lines are the Landaus levels without strain and with strain $\alpha = 0.3$ for $B = 1$~eV$^2$}
		\label{fig.cont}
	\end{figure}

	It is crucial to note that the nature of eigen-function allows the quantites like expectation value of a current density $\langle J \rangle$ to take a non zero value giving rise to magneto-optical conductivity~\cite{li.carbotte.13,zhao.sun.22}.
	This nature of eigen-functions does not gets affected with an external strain on the system and hence we expect more or less similar magneto-optical response from WSM under strain.

	As evident from Fig.~\ref{fig.cont} due to the application of external strain the Landau levels gets shifted with the preserved dependence on $k_z$. These new shifted Landau levels can be explained either with a renormalised magnetic field, $B \rightarrow (1 - \alpha^2) B$ or a renormalised hopping parameter $t \rightarrow (1 - \alpha^2) t$. Even though both ways lead to same result former relies on changing the magnetic field which keeps the underlying system unchanged. Later changes the hopping parameter which affects the crystal itself leading to change in system. Introduction of strain gives us very subtle insight and facilitates study of multiple hopping parameter by adjusting the magnetic field.

Moreover, the zeroth Landau levels (conduction and valence bands) still have the degeneracies. The energy of the strained system shows the strongest effect of strain in the vicinity of either Weyl node i.e. $k_z \simeq \pm \sqrt{\Delta}$ for and beyond first LL. It can seen from Eq.~(\ref{eq:energy}) that the effect strain has on energy depends on the chirality of node.
	We now direct our attention to an alternate model of WSM which lacks the cylindrical symmetry which Eq.~(\ref{eq.ham_cont}) shows in $k_x - k_y$ plane. Such a WSM will have different hopping along $x$ (say $t_x$), and $y$ (say $t_y$). This asymmetric WSM is equivalent to Eq.~(\ref{eq.ham_str}) this makes the strained Hamiltonian a bridge between the symmetric and asymmetric WSM, and allows us to study both simultaneously.


	\begin{figure}[!hb]
		\centering
		\includegraphics[width=0.99\linewidth]{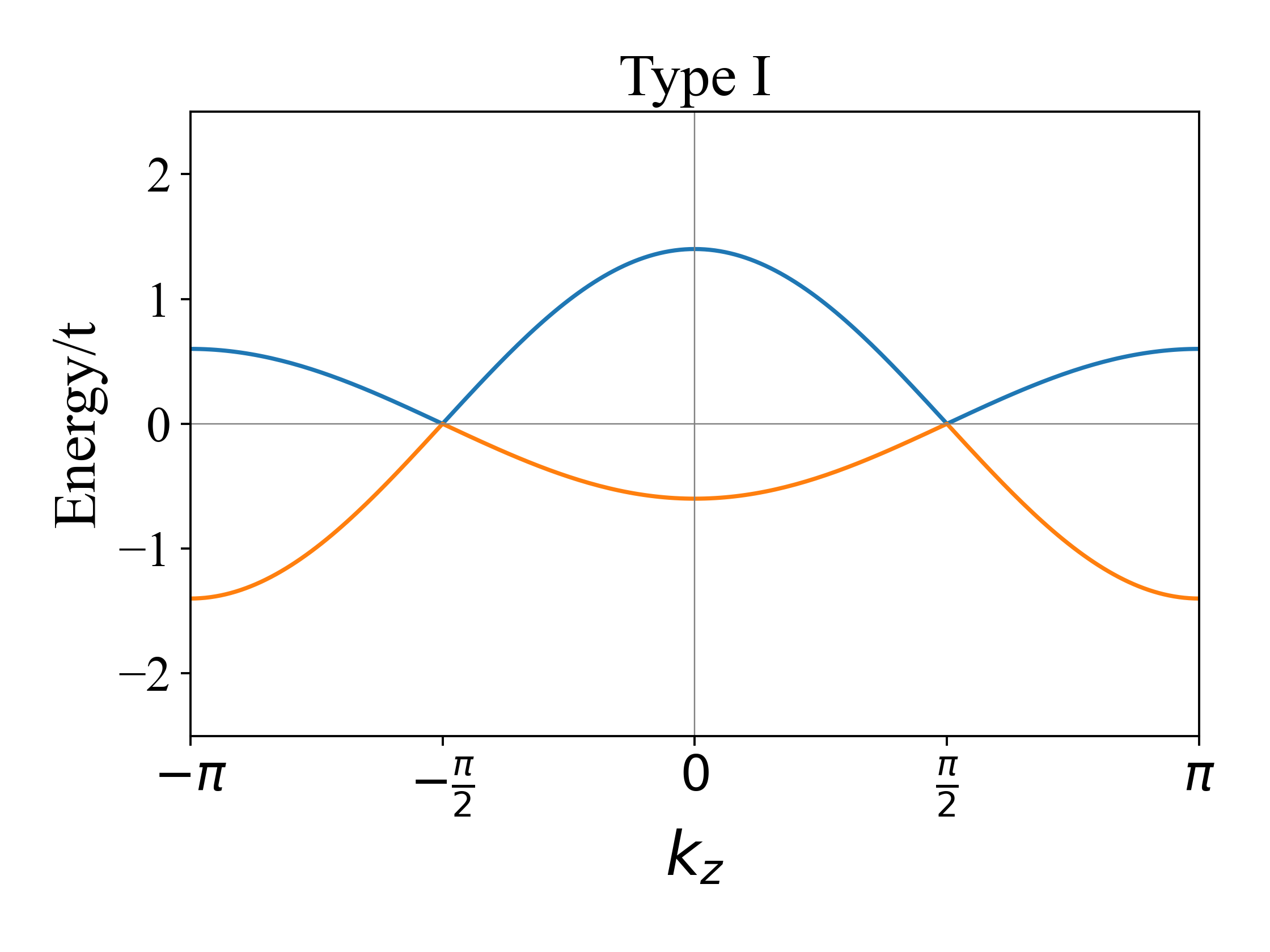}\\
		\vspace*{0.5cm}
		\includegraphics[width=0.99\linewidth]{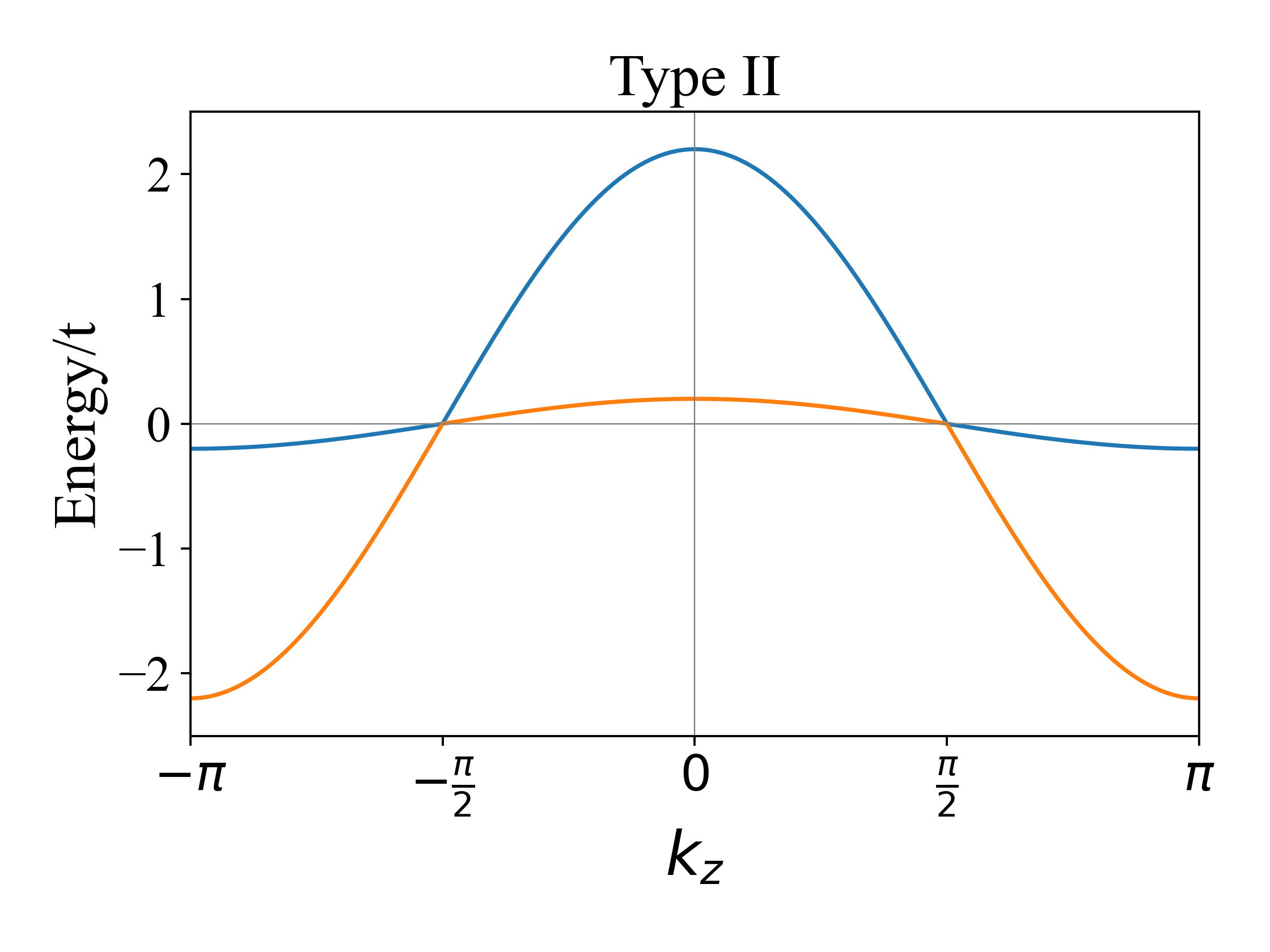}
		\caption{
			The band structure along $k_z$ (for $k_x = k_y = 0$) for the discussed lattice model in the case of type-I and type-II Weyl nodes (left and right panel, respectively).
			Tilt factor is $t_0 / t = 0.4$ for type-I, and $t_0/t = 1.2$ for type-II.
			\label{fig.tightbinding}
		}
	\end{figure}

	\section{Lattice model description}
	\label{sec.latt}

	To illustrate the role of the interplay between uniaxial strain and external magnetic field, we investigate the two-band lattice model describing the type-I and type-II WSM~\cite{roy.goswami.17}.
	In this case, the system is described by the Hamiltonian~\cite{nag.menon.20}:
	\begin{eqnarray}
		H =\pmb{N}_{k} \cdot \pmb{\sigma}
	\end{eqnarray}
	Here $\pmb{\sigma} = ( \sigma_{0}, \sigma_{x}, \sigma_{y}, \sigma_{z} )$ denotes the Pauli matrix vector, while $\pmb{N}_{k}$ is the momentum-dependent form factor, which takes the form~\footnote{Details of the model can be found in Supplemental Material of Ref.~\cite{nag.menon.20}.}: $N_{0} = t_{0} ( \cos k_{z} + \cos k_{x} - 1)$, $N_{x} = t \sin k_{x}$, $N_{y} = t \sin k_{y}$, and $N_{z} = t_{z} \cos k_{z} - m_{z} + t_{0} ( 2 - \cos k_{x} - \cos k_{y} )$.
	In this model, the Weyl nodes are located at ${\bm k} = ( 0, 0, \pm k_{0} )$ with:
	\begin{eqnarray}
		t_{z} \cos k_{0} = m_{z} + t_{0} \cos k_{x} + t_{0} \cos k_{y} - 2 t_{0} .
	\end{eqnarray}
	For $m_{z} = 0$, the Weyl nodes are at $k_{0} = \pm \pi / 2$.
	Presented model is an extension of the model presented in Ref.~\cite{nag.nandy.20}, and described type-I and type-II WSM for $t_{0} < t_{z}$ and $t_{0} \geq t_{z} $, respectively.
	These correspond to the tilted or overtilted Weyl node (type-I and type-II, respectively) and example of the energy dispersion along $k_z$ around the Weyl nodes is depicted in Fig.~\ref{fig.tightbinding} for both type-I ($t_0 / t = 0.4$) and type-II ($t_0 / t = 1.2$).

	In the real space the presented model take the form:
	\begin{widetext}
		\begin{eqnarray}
			\nonumber H &=& \sum_{i s} \left[\frac{t_0}{2} \left( c_{i \pm \hat{x}, s}^{\dagger} c_{i s} + c_{i \pm \hat{z}, s}^{\dagger} c_{i s} \right) - t_0 c_{i s}^{\dagger} c_{i s} \right] + \sum_{i s s'}  \sigma_{s s'}^z \left[ - \frac{t_0}{2} \left( c_{i \pm \hat{x}, s}^{\dagger} c_{i s'} + c_{i \pm \hat{y}, s}^{\dagger} c_{i s'} \right) + \frac{t_z}{2} c_{i \pm \hat{z} s}^{\dagger} c_{i s'}  - \tilde{\mu} c_{i s}^{\dagger} c_{i s} \delta_{s s'} \right] \\
			\label{eq.real_space} &+& \sum_{i s s'} \left( \frac{t}{2} c_{i + \hat{x}, s}^{\dagger} \sigma^{y}_{s s'}  c_{i s'} 
			- \imath \frac{t}{2} c_{i + \hat{y}, s}^{\dagger} \sigma^{y}_{s s'} c_{i s'} + H.c.\right)
		\end{eqnarray}
	\end{widetext}
	where $\hat{e} = \{ \hat{x}, \hat{y}, \hat{z} \}$ denotes unit vectors along $x$, $y$, and $z$ direction, while $\tilde{\mu} = m_z - 2 t_0$.
	Here $c_{i s}^{\dagger}$ ($c_{i s}$) denotes the creation (annihilation) operator of electron in site $i$ with pseudospin $s$.
	As we can see, the first line correspond to the ``free'' electron term (i.e. hopping between neighboring sites and effective on-site terms), while second line correspond to the spin--orbit coupling.

	We plan to determine the energy spectrum of WSM with a cubic lattice crystal with respect to magnetic field.
	In order to accomplish this we introduce the magnetic field using the Pierel's substitution~\cite{peierls.33}:
	\begin{eqnarray}
		t_{ij} \left( \mathbf{A} \right) = t \exp \left( \imath \frac{2 \pi}{\phi_{0}} \int_{\mathbf{R}_{i}}^{\mathbf{R}_{j}} \mathbf{A} \cdot d\mathbf{r} \right)
	\end{eqnarray}
	applied for hopping integrals in kinetic part of the real space Hamiltonian~(\ref{eq.real_space}).
	For simplification, but without lost generality, we assume Landau gauge $\mathbf{A} = (0, Bx, 0)$, which gives $\mathbf{B} = \text{rot} \; \mathbf{A} \parallel \hat{z}$ (this corresponds directly to the situation discussed in previous Section).
	The real space Hamiltonian can be derived by inverse Fourier transform of the Hamiltonian by using the following substitution:
	\begin{eqnarray}
		c_{x {\bm k}' s}^{\dagger} &=& \frac{1}{ \sqrt{N} } \sum_{i} c_{i s}^{\dagger} e^{-\imath {\bm k}' \cdot {\bm r}_{i}} .
	\end{eqnarray}
	Here, ${\bm r}_{i}$ denotes position of site $i$ in $yz$ plane.
	As consequence coordinates along $x$ are unchanged.
	This transformation act only in $yz$ plane, i.e. changed only two out of three space coordinates: $\left( x, y, z \right) \longrightarrow \left( x, k_{y}, k_{z} \right)$.
	After this derivation, the Hamiltonian~(\ref{eq.real_space}) takes the well known form of the Harper equation~\cite{harper.55}:
	\begin{widetext}
		\begin{eqnarray}
			H &=& \sum_{x {\bm k}' s s'} \sigma_{s s'}^z \left[ - \frac{t_0}{2} c_{x \pm 1, {\bm k}' s}^{\dagger} c_{x {\bm k}' s'} + \left(t_z \cos k_{z} - t_0 \cos \left( k_{y} + 2 \pi \frac{\phi x}{\phi_{0}}  \right) - \tilde{\mu} \right) c_{x {\bm k}' s}^{\dagger} c_{x {\bm k}' s'} \right]  \\
			&+& \sum_{x {\bm k}' s} \left[ t_0 \, \cos k_z \, c_{x {\bm k}' s}^{\dagger} c_{x {\bm k}' s} - t_0 \, c_{x {\bm k}' s}^{\dagger} c_{x {\bm k}' s} + \frac{t_0}{2} c_{x \pm 1, {\bm k}' s}^{\dagger} c_{x {\bm k}' s}  \right] + \sum_{x {\bm k}' s s'} \sigma_{s s'}^y \left[  \frac{t}{2} c_{x \pm 1, {\bm k}' s}^{\dagger} c_{x {\bm k}' s'} + t \sin k_{y} c_{x {\bm k}' s}^{\dagger} c_{x {\bm k}' s'} \right] \nonumber
		\end{eqnarray}
	\end{widetext}
	where $\phi_{0}$ is the quantum of magnetic flux.

	We note that the pseudo-spin exchange parameter does not gets affected by the magnetic field. 
	To build the energy spectrum with respect to the magnetic field $\phi$, we consider a crystal for which we have $N$ lattice sites along $x$ direction. 
	The energy spectrum can be studied only for certain values of magnetic field and it must satisfy $2 \pi \phi x = \mathbb{Z}^{+} \phi_{0}$, where $\mathbb{Z}^+$ are positive integers.

	As before we did in Sec.~\ref{sec.cont}, we reintroduce the uniaxial strain by rescaling the hopping integrals along $x$ and $y$ directions:
	\begin{eqnarray}
		t_{x} \rightarrow \gamma_{-} t_{x} , \quad \text{and} \quad t_{y} \rightarrow \gamma_{+} t_{y} ,
	\end{eqnarray}
	where $t_{i}$ denotes the hopping integrals along $i$ direction in Eq.~(\ref{eq.real_space}), while $\gamma_{\pm}$ depends by uniaxial strain $\alpha$, and is defined in Eq.~(\ref{eq.train_k}).

	\begin{figure}[!b]
		\centering
		\includegraphics[width=\linewidth]{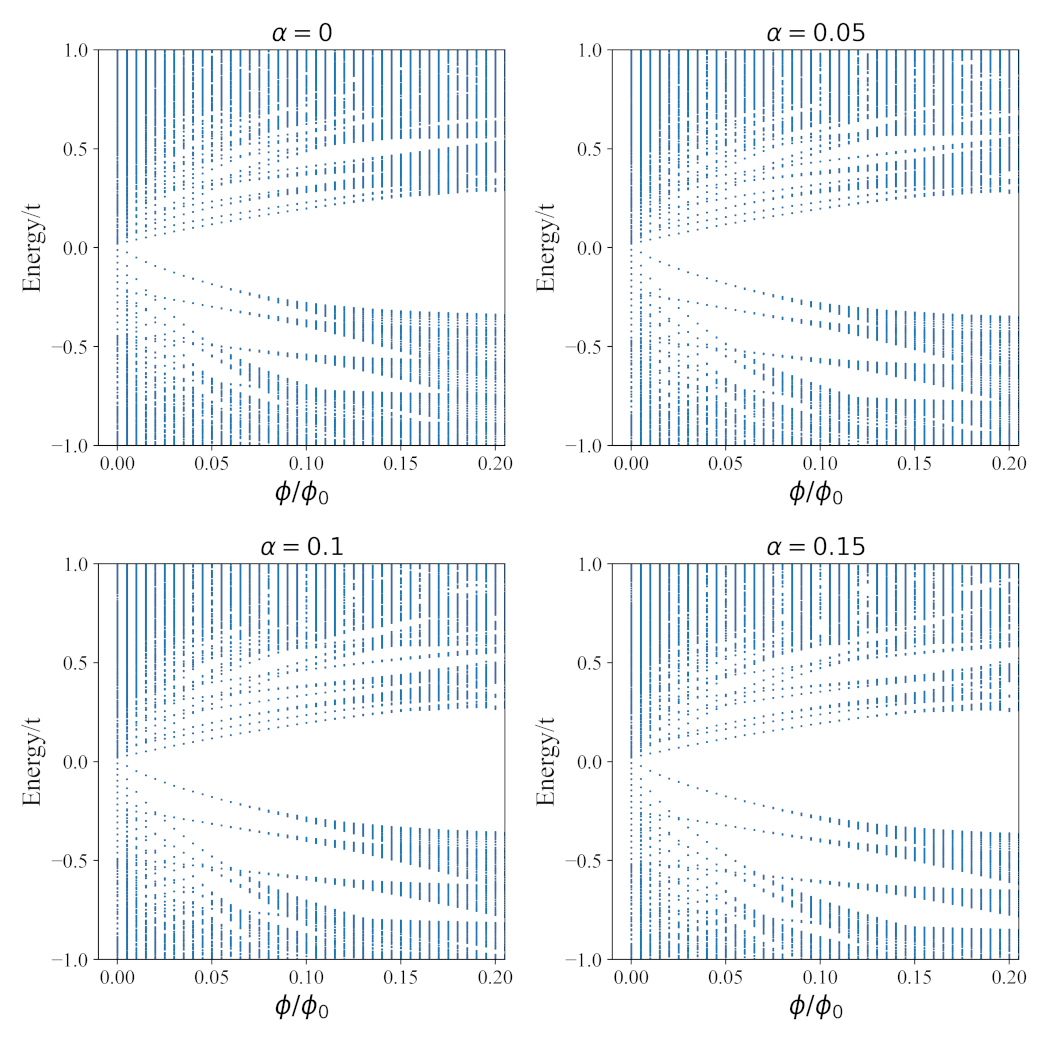}
		\caption{
			Comparison of the energy spectrum for type-I lattice model for several values of the uniaxial strain $\alpha$ as a function of magnetic flux $\phi$.
			Results for $t_0 / t = 0.4$ eV, and $t_z / t = 1$ in $200 \times 200 \times 13$ lattice, for $k_z$ closest to the Weyl node.
			\label{fig.strain}
		}
	\end{figure}

	\begin{figure}[!b]
		\centering
		\includegraphics[width=\linewidth]{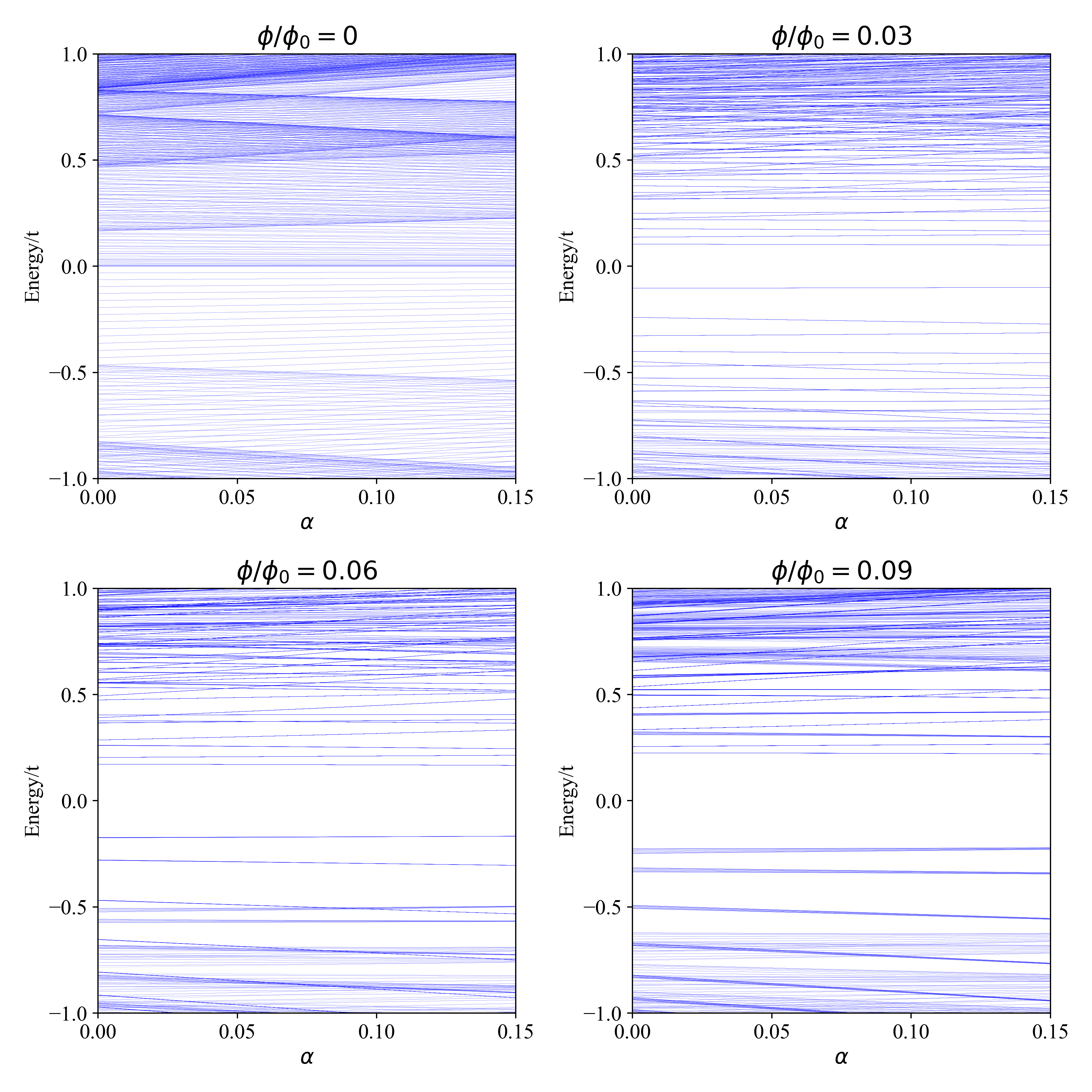}
		\caption{
			Comparison of the energy spectrum for type-I lattice model for several magnetic flux $\phi$ as function of uniaxial strain $\alpha$.
			Results for $t_0 / t = 0.4$ eV, in $200 \times 200 \times 13$ lattice, for $k_z$ closest to the Weyl node.
			\label{fig.strain1}
		}
	\end{figure}

	\subsection{System spectrum}

	The energy dispersion of the discussed lattice model in the absence of magnetic field hosts Weyl nodes at ${\bm k} = \left( 0, 0, \pm \pi/2 \right)$. 
	Tilt of the Weyl nodes is controlled by the ratio $t_{0} / t_{z}$.
	Modification of this ratio allows the realization of type-I and type-II Weyl nodes.
	In our calculation we set $t_{0} = 0.4 \, t_{z}$ for type-I model, and $t_{0} = 1.2 \, t_{z}$ for type-II.
	Additionally the parameter $m_z$ controls the energy level at which Weyl nodes are formed.
	For simplification, $m_{z}$ has been set to zero which leads to the Weyl nodes to form at the Fermi level (``zero'' energy level).

	We start the discussion with the role of the magnetic flux $\phi$ in the presence of fixed uniaxial strain $\alpha$ (Fig.~\ref{fig.strain}). 
	As we apply a magnetic field, the effect of strain on LLs begins to emerge.
	Each of these spectra resemble the famous Hofstadter butterfly~\cite{hofstadter.76}.
	Applying a unaxial strain $\alpha$ lead to the renormalization of the system spectra (cf.~all panels).
	In practice, the main features of the system spectra are very similar (e.g. shape of gaps between energy levels).
	As for typical Hofstadter butterfly, the presented spectra possess the fractal feature. 
	Nevertheless, we reiterate that the effect of strain is considered within linear response and the only small values of $\alpha$ correspond to the realistic (physical) conditions.

	For small strain $\alpha$ we observe the spectrum renormalization in the magnetic flux $\phi$ domain.
	In relatively simpler way, we can recognize the increase of the gaps between the LLs.
	Additional shift of the energy level can be observed by comparing the spectra for two different strains $\alpha$ [e.g.~Fig.~SF1 in the Supplemental Material (SM)~\footnote{See Supplemental Material at [URL will be inserted by publisher] for additional theoretical results.}].
	As we mention in the previous Section, fixed $\alpha \neq 0$ leads to the magnetic field renormalization.
	Indeed, the energy spectra for $\phi ( \alpha \neq 0 )$ can be mapped to spectra of $\phi' ( \alpha = 0 )$ which corresponds to renormalization of magnetic flux $\phi$.
Nevertheless, this feature does not hold for LLs far away from the Fermi level i.e. the Weyl nodes.

	Similar analyses can be performed for the fixed magnetic flux $\phi$ as a function of the uniaxial strain $\alpha$ (Fig.~\ref{fig.strain1}).
	In the absence of magnetic flux ($\phi = 0$) the strain creates a trend where the bands bend away from the Fermi level.
	Similar to the previous case, we see formation of the LLs (cf.~all bands).
	In reasonable range of $\alpha$ the LLs are unchanged.
	Some of them bends away from the Fermi level and some towards the Fermi level.

	In order to study the effects of magnetic field on Weyl nodes we only show the energy spectra very close to $k_z = \pm \pi/2$. 
	The spectra is made up of superimposed fractals each of which correspond to a momenta along $k_y$.  We can observe that for both type-I and type-II the energy spectra near the zero energy and small magnetic field bifurcates into multiple LLs. 
	These levels then get intertwined to give the spectra its fractal nature. 
	Increasing the number of lattice sties along the $x$ direction only adds details to the fractal nature of the spectra. 
	Increase in the number of lattice sites along any other direction ends up being a stacking of fractals for other momentum values. 
	The fractal nature is evident in the energy spectra and density of states with respect to the magnetic (see Fig.~SF2 in the SM~\cite{Note1}).
	The scale of energy range in type-I WSM leads to merging of the spectra corresponding to different values of $k_y$. 
	Type-II on the other hand has a contrasting nature of spectra, owing to its larger energy range. 
	Not only type-II spectra holds the huge gaps in the valence band the stacking does not obscures the fractal features of spectra when compared to the type-I. 
	Both density of states and energy spectra have a clear separation between the valance and conduction part. 
	The valence part of the fractal is a transformation of the valance part which can be checked by inverting the sign of tilt. 
	This inversion of tilt produces a result which is flipped about Fermi level. 
	
	Moreover, in Fig.~SF3 in the SM~\cite{Note1}, we display the energy dispersion for type-I Weyl node along $k_z$ for $k_y = 0$ which is the closest to the Weyl node. All we can notice is that the introduction of the magnetic field leads to formation of LLs. 
	With the increasing magnetic field the LLs become more sharp. Applying uniaxial strain introduces very tiny changes in the system, this is again behaves as scaling.

	\subsection{Role of the Weyl node tilted}

	We discuss the lattice model which describes both type-I and type-II Weyl nodes.
	Here, the type of Weyl node is controlled by $t_{0}$.
	Thus, by modifying the parameter $t_{0}$ we can have a  continuous change from Weyl node of type-I to type-II (for $t_{0} < t_{z}$ and $t_{0} \geq t_{z}$, respectively).
	For the critical value $t_{0} = t_{z}$ (called also type-III~\cite{huang.jin.18,liu.sun.18,fragkos.tsipas.21}), the the Weyl node possess the flat band along one direction.

	Spectra of the system with a varying Weyl node tilt (i.e. as a function of $t_0$) is presented in Fig.~SF4 in the SM~\cite{Note1}. The figures depict the energy spectrum for a slice of WSM with a finite size of $L_x \times L_y \times L_z = 100 \times 13 \times 100$. The slice chosen is such that $k_y = 0$ and $k_z = \pi/2$ hence sampling the Weyl node for which the tilt is equal to the tilt $t_0/t_z$ [c.f. Fig.~\ref{fig.tightbinding}]. The figures in first row show the energy levels for all possible $k_x$ in the absence of magnetic field. These figures demonstrate how strain affects the energy for all tilt angles. It also crucial to note that although it is possible to get an analogous result with continuum model without magnetic field for a certain momenta, the finite size effects seen here are unique to the lattice model. 
	The structure of LLs for fixed magnetic flux $\phi$ is preserved for changing uniaxial strain $\alpha$ (cf.~panels in the same row).
	Nevertheless, it is apparent that for fixed magnetic field $\phi$, increasing uniaxial strain $\alpha$ resembles scaling of the energy. This is further illustrated in the Fig.~SF3 in the SM~\cite{Note1}, the energy is rescaled for all magnetic fields (across rows) as we increase the strain (along columns). The fact that energy rescales by varying the strain further exemplifies the renormalization of either the hopping parameter or the magnetic field. Aided with the Eq.~(\ref{eq.mag_len}) we can say that in the presence of magnetic field the system perceives a normalised magnetic field. On the other hand if the WSM under consideration is totally asymmetric ($t_x \neq t_y$) we can find unique $t$ and $\alpha$ such that 
	\begin{eqnarray}
		t_x = t \, (1 - \alpha) \quad \text{and} \quad t_y = t \, (1 + \alpha) . \nonumber
	\end{eqnarray}
	As the tilt of the cone increases the absolute value of slope of the LLs also increases this causes the bunching of LLs as tilt increases. Additionally, increasing $\phi$ for fixed $\alpha$ alters the LLs  and the tilt causes them to merge forming tight bands which moves with tilt $t_0$. The asymmetry about zero energy in Fig.~SF4 in the SM~\cite{Note1} arises due to the tilt which introduces a lopsided nature in the band structure of the tight binding Hamiltonian along $k_y = 0$ and $k_z = \pi/2$. If both nodes are taken into account (which would imply plotting the entire spectrum) then the asymmetry does not persists.
	We should also notice that the energy seamlessly changes around $t_{0} / t_{z} = 1$.


	\section{Summary}
	\label{sec.sum}

	In this paper we present study of the Weyl node in the presence of uniaxial strain $\alpha$.
	In the continuum model, we observe that the Landau levels appear with their values being tuned by the magnetic length $l_{B} \sim 1 / \sqrt{1-\alpha^{2}}$.
	This lead to the shift of Landau levels, and introduces assymmetry in the system. 
	For small uniaxial strain, the energy behaves quadratically with $\alpha$.
	Observing the magnetic length and choosing to renormalize either $t$ or $B$ makes introduction of strain an excellent tool to study not only a strained Weyl node but also study completely assymmetric Weyl node all at once.
	These calculations can be carried over to a tilted Weyl node by applying a non-unitary boost to the system.

	Similarly, we present numerical study of the lattice model of type-I and type-II Weyl nodes.
	Uniaxial strain was introduced by rescaling the hopping integrals, while the magnetic field via the Pierels substitution.
	As can be expected, applying the magnetic field leads to a fractal-like spectra, which resembles the Hofstadter's butterfly. 
	The energy spectrum show distinct features for type-I and type-II Weyl nodes.
	Introduction of strain changes the density of states by changing the alignment of the spectra for momenta along $k_y$. 
	Our study of tight binding model shows that it is reasonable to renormalize the magnetic field for small values of strain.
	
		In summary, using the continuum model we show that the uniaxial strain affects the lowest LLs of the spectrum energy relatively weakly.
		The same conclusion were drawn using the lattice model with type-I and type-II Weyl nodes (with modified hopping integrals due to the strain and Pierel's substitution associated with external magnetic field).
		This results is in agreement with obtained previously~\cite{pikulin.chen.16}, where the strain weakly modified energy spectrum.
		In the case of lattice model, for fixed uniaxal strain the magnetic field can strongly modify the energy spectrum creating the Hofstadter butterfly structure.
		This feature was not reported earlier under uniaxial strain.

	\begin{acknowledgments}
		S.Y. and A.P. are grateful to Laboratoire de Physique des Solides in Orsay (CNRS, University Paris Saclay) for hospitality during a part of the work on this project.
		S.Y. acknowledges financial support provided by the Polish National Agency for Academic Exchange NAWA  under the Programme STER-- Internationalisation of doctoral schools, Project no.~PPI/STE/2020/1/00020.
		We kindly acknowledge support by National Science Centre (NCN, Poland) under Project No.~2021/43/B/ST3/02166.
	\end{acknowledgments}

	\bibliography{biblio.bib}

\begin{thebibliography}{48}%
\makeatletter
\providecommand \@ifxundefined [1]{%
 \@ifx{#1\undefined}
}%
\providecommand \@ifnum [1]{%
 \ifnum #1\expandafter \@firstoftwo
 \else \expandafter \@secondoftwo
 \fi
}%
\providecommand \@ifx [1]{%
 \ifx #1\expandafter \@firstoftwo
 \else \expandafter \@secondoftwo
 \fi
}%
\providecommand \natexlab [1]{#1}%
\providecommand \enquote  [1]{``#1''}%
\providecommand \bibnamefont  [1]{#1}%
\providecommand \bibfnamefont [1]{#1}%
\providecommand \citenamefont [1]{#1}%
\providecommand \href@noop [0]{\@secondoftwo}%
\providecommand \href [0]{\begingroup \@sanitize@url \@href}%
\providecommand \@href[1]{\@@startlink{#1}\@@href}%
\providecommand \@@href[1]{\endgroup#1\@@endlink}%
\providecommand \@sanitize@url [0]{\catcode `\\12\catcode `\$12\catcode
  `\&12\catcode `\#12\catcode `\^12\catcode `\_12\catcode `\%12\relax}%
\providecommand \@@startlink[1]{}%
\providecommand \@@endlink[0]{}%
\providecommand \url  [0]{\begingroup\@sanitize@url \@url }%
\providecommand \@url [1]{\endgroup\@href {#1}{\urlprefix }}%
\providecommand \urlprefix  [0]{URL }%
\providecommand \Eprint [0]{\href }%
\providecommand \doibase [0]{https://doi.org/}%
\providecommand \selectlanguage [0]{\@gobble}%
\providecommand \bibinfo  [0]{\@secondoftwo}%
\providecommand \bibfield  [0]{\@secondoftwo}%
\providecommand \translation [1]{[#1]}%
\providecommand \BibitemOpen [0]{}%
\providecommand \bibitemStop [0]{}%
\providecommand \bibitemNoStop [0]{.\EOS\space}%
\providecommand \EOS [0]{\spacefactor3000\relax}%
\providecommand \BibitemShut  [1]{\csname bibitem#1\endcsname}%
\let\auto@bib@innerbib\@empty
\bibitem [{\citenamefont {Guinea}\ \emph {et~al.}(2010)\citenamefont {Guinea},
  \citenamefont {Katsnelson},\ and\ \citenamefont
  {Geim}}]{guinea.katsnelson.10}%
  \BibitemOpen
  \bibfield  {author} {\bibinfo {author} {\bibfnamefont {F.}~\bibnamefont
  {Guinea}}, \bibinfo {author} {\bibfnamefont {M.~I.}\ \bibnamefont
  {Katsnelson}},\ and\ \bibinfo {author} {\bibfnamefont {A.~K.}\ \bibnamefont
  {Geim}},\ }\bibfield  {title} {\bibinfo {title} {Energy gaps and a zero-field
  quantum {Hall} effect in graphene by strain engineering},\ }\href
  {https://doi.org/10.1038/nphys1420} {\bibfield  {journal} {\bibinfo
  {journal} {Nat. Phys.}\ }\textbf {\bibinfo {volume} {6}},\ \bibinfo {pages}
  {30} (\bibinfo {year} {2010})}\BibitemShut {NoStop}%
\bibitem [{\citenamefont {Levy}\ \emph {et~al.}(2010)\citenamefont {Levy},
  \citenamefont {Burke}, \citenamefont {Meaker}, \citenamefont {Panlasigui},
  \citenamefont {Zettl}, \citenamefont {Guinea}, \citenamefont {Neto},\ and\
  \citenamefont {Crommie}}]{levy.burke.10}%
  \BibitemOpen
  \bibfield  {author} {\bibinfo {author} {\bibfnamefont {N.}~\bibnamefont
  {Levy}}, \bibinfo {author} {\bibfnamefont {S.~A.}\ \bibnamefont {Burke}},
  \bibinfo {author} {\bibfnamefont {K.~L.}\ \bibnamefont {Meaker}}, \bibinfo
  {author} {\bibfnamefont {M.}~\bibnamefont {Panlasigui}}, \bibinfo {author}
  {\bibfnamefont {A.}~\bibnamefont {Zettl}}, \bibinfo {author} {\bibfnamefont
  {F.}~\bibnamefont {Guinea}}, \bibinfo {author} {\bibfnamefont {A.~H.~C.}\
  \bibnamefont {Neto}},\ and\ \bibinfo {author} {\bibfnamefont {M.~F.}\
  \bibnamefont {Crommie}},\ }\bibfield  {title} {\bibinfo {title}
  {Strain-induced pseudo–magnetic fields greater than 300 tesla in graphene
  nanobubbles},\ }\href {https://doi.org/10.1126/science.1191700} {\bibfield
  {journal} {\bibinfo  {journal} {Science}\ }\textbf {\bibinfo {volume}
  {329}},\ \bibinfo {pages} {544} (\bibinfo {year} {2010})}\BibitemShut
  {NoStop}%
\bibitem [{\citenamefont {Tchoumakov}\ \emph {et~al.}(2016)\citenamefont
  {Tchoumakov}, \citenamefont {Civelli},\ and\ \citenamefont
  {Goerbig}}]{tchoumakov.civelli.16}%
  \BibitemOpen
  \bibfield  {author} {\bibinfo {author} {\bibfnamefont {S.}~\bibnamefont
  {Tchoumakov}}, \bibinfo {author} {\bibfnamefont {M.}~\bibnamefont
  {Civelli}},\ and\ \bibinfo {author} {\bibfnamefont {M.~O.}\ \bibnamefont
  {Goerbig}},\ }\bibfield  {title} {\bibinfo {title} {Magnetic-field-induced
  relativistic properties in type-{I} and type-{II} {Weyl} semimetals},\ }\href
  {https://doi.org/10.1103/PhysRevLett.117.086402} {\bibfield  {journal}
  {\bibinfo  {journal} {Phys. Rev. Lett.}\ }\textbf {\bibinfo {volume} {117}},\
  \bibinfo {pages} {086402} (\bibinfo {year} {2016})}\BibitemShut {NoStop}%
\bibitem [{\citenamefont {Pikulin}\ \emph {et~al.}(2016)\citenamefont
  {Pikulin}, \citenamefont {Chen},\ and\ \citenamefont
  {Franz}}]{pikulin.chen.16}%
  \BibitemOpen
  \bibfield  {author} {\bibinfo {author} {\bibfnamefont {D.~I.}\ \bibnamefont
  {Pikulin}}, \bibinfo {author} {\bibfnamefont {A.}~\bibnamefont {Chen}},\ and\
  \bibinfo {author} {\bibfnamefont {M.}~\bibnamefont {Franz}},\ }\bibfield
  {title} {\bibinfo {title} {Chiral anomaly from strain-induced gauge fields in
  dirac and weyl semimetals},\ }\href
  {https://doi.org/10.1103/PhysRevX.6.041021} {\bibfield  {journal} {\bibinfo
  {journal} {Phys. Rev. X}\ }\textbf {\bibinfo {volume} {6}},\ \bibinfo {pages}
  {041021} (\bibinfo {year} {2016})}\BibitemShut {NoStop}%
\bibitem [{\citenamefont {Shapourian}\ \emph {et~al.}(2015)\citenamefont
  {Shapourian}, \citenamefont {Hughes},\ and\ \citenamefont
  {Ryu}}]{shapourian.hughes.15}%
  \BibitemOpen
  \bibfield  {author} {\bibinfo {author} {\bibfnamefont {H.}~\bibnamefont
  {Shapourian}}, \bibinfo {author} {\bibfnamefont {T.~L.}\ \bibnamefont
  {Hughes}},\ and\ \bibinfo {author} {\bibfnamefont {S.}~\bibnamefont {Ryu}},\
  }\bibfield  {title} {\bibinfo {title} {Viscoelastic response of topological
  tight-binding models in two and three dimensions},\ }\href
  {https://doi.org/10.1103/PhysRevB.92.165131} {\bibfield  {journal} {\bibinfo
  {journal} {Phys. Rev. B}\ }\textbf {\bibinfo {volume} {92}},\ \bibinfo
  {pages} {165131} (\bibinfo {year} {2015})}\BibitemShut {NoStop}%
\bibitem [{\citenamefont {Cortijo}\ \emph {et~al.}(2015)\citenamefont
  {Cortijo}, \citenamefont {Ferreir\'os}, \citenamefont {Landsteiner},\ and\
  \citenamefont {Vozmediano}}]{cortijo.ferreiros.15}%
  \BibitemOpen
  \bibfield  {author} {\bibinfo {author} {\bibfnamefont {A.}~\bibnamefont
  {Cortijo}}, \bibinfo {author} {\bibfnamefont {Y.}~\bibnamefont
  {Ferreir\'os}}, \bibinfo {author} {\bibfnamefont {K.}~\bibnamefont
  {Landsteiner}},\ and\ \bibinfo {author} {\bibfnamefont {M.~A.~H.}\
  \bibnamefont {Vozmediano}},\ }\bibfield  {title} {\bibinfo {title} {Elastic
  gauge fields in {Weyl} semimetals},\ }\href
  {https://doi.org/10.1103/PhysRevLett.115.177202} {\bibfield  {journal}
  {\bibinfo  {journal} {Phys. Rev. Lett.}\ }\textbf {\bibinfo {volume} {115}},\
  \bibinfo {pages} {177202} (\bibinfo {year} {2015})}\BibitemShut {NoStop}%
\bibitem [{\citenamefont {Vazifeh}\ and\ \citenamefont
  {Franz}(2013)}]{vazifeh.franz.13}%
  \BibitemOpen
  \bibfield  {author} {\bibinfo {author} {\bibfnamefont {M.~M.}\ \bibnamefont
  {Vazifeh}}\ and\ \bibinfo {author} {\bibfnamefont {M.}~\bibnamefont
  {Franz}},\ }\bibfield  {title} {\bibinfo {title} {Electromagnetic response of
  {Weyl} semimetals},\ }\href {https://doi.org/10.1103/PhysRevLett.111.027201}
  {\bibfield  {journal} {\bibinfo  {journal} {Phys. Rev. Lett.}\ }\textbf
  {\bibinfo {volume} {111}},\ \bibinfo {pages} {027201} (\bibinfo {year}
  {2013})}\BibitemShut {NoStop}%
\bibitem [{\citenamefont {Lee}\ \emph {et~al.}(2022)\citenamefont {Lee},
  \citenamefont {Park},\ and\ \citenamefont {Vozmediano}}]{lee.park.22}%
  \BibitemOpen
  \bibfield  {author} {\bibinfo {author} {\bibfnamefont {Y.-J.}\ \bibnamefont
  {Lee}}, \bibinfo {author} {\bibfnamefont {C.-H.}\ \bibnamefont {Park}},\ and\
  \bibinfo {author} {\bibfnamefont {M.~A.~H.}\ \bibnamefont {Vozmediano}},\
  }\bibfield  {title} {\bibinfo {title} {Strain-induced collapse of {Landau}
  levels in real {Weyl} semimetals},\ }\href
  {https://doi.org/10.1103/PhysRevB.106.075125} {\bibfield  {journal} {\bibinfo
   {journal} {Phys. Rev. B}\ }\textbf {\bibinfo {volume} {106}},\ \bibinfo
  {pages} {075125} (\bibinfo {year} {2022})}\BibitemShut {NoStop}%
\bibitem [{\citenamefont {Arjona}\ \emph {et~al.}(2017)\citenamefont {Arjona},
  \citenamefont {Castro},\ and\ \citenamefont {Vozmediano}}]{arjona.castro.17}%
  \BibitemOpen
  \bibfield  {author} {\bibinfo {author} {\bibfnamefont {V.}~\bibnamefont
  {Arjona}}, \bibinfo {author} {\bibfnamefont {E.~V.}\ \bibnamefont {Castro}},\
  and\ \bibinfo {author} {\bibfnamefont {M.~A.~H.}\ \bibnamefont
  {Vozmediano}},\ }\bibfield  {title} {\bibinfo {title} {Collapse of {Landau}
  levels in {Weyl} semimetals},\ }\href
  {https://doi.org/10.1103/PhysRevB.96.081110} {\bibfield  {journal} {\bibinfo
  {journal} {Phys. Rev. B}\ }\textbf {\bibinfo {volume} {96}},\ \bibinfo
  {pages} {081110} (\bibinfo {year} {2017})}\BibitemShut {NoStop}%
\bibitem [{\citenamefont {Yan}\ and\ \citenamefont
  {Felser}(2017)}]{yan.felser.17}%
  \BibitemOpen
  \bibfield  {author} {\bibinfo {author} {\bibfnamefont {B.}~\bibnamefont
  {Yan}}\ and\ \bibinfo {author} {\bibfnamefont {C.}~\bibnamefont {Felser}},\
  }\bibfield  {title} {\bibinfo {title} {Topological materials: Weyl
  semimetals},\ }\href
  {https://doi.org/10.1146/annurev-conmatphys-031016-025458} {\bibfield
  {journal} {\bibinfo  {journal} {Annu. Rev. Condens. Matter Phys.}\ }\textbf
  {\bibinfo {volume} {8}},\ \bibinfo {pages} {337} (\bibinfo {year}
  {2017})}\BibitemShut {NoStop}%
\bibitem [{\citenamefont {Armitage}\ \emph {et~al.}(2018)\citenamefont
  {Armitage}, \citenamefont {Mele},\ and\ \citenamefont
  {Vishwanath}}]{armitage.mele.18}%
  \BibitemOpen
  \bibfield  {author} {\bibinfo {author} {\bibfnamefont {N.~P.}\ \bibnamefont
  {Armitage}}, \bibinfo {author} {\bibfnamefont {E.~J.}\ \bibnamefont {Mele}},\
  and\ \bibinfo {author} {\bibfnamefont {A.}~\bibnamefont {Vishwanath}},\
  }\bibfield  {title} {\bibinfo {title} {Weyl and {Dirac} semimetals in
  three-dimensional solids},\ }\href
  {https://doi.org/10.1103/RevModPhys.90.015001} {\bibfield  {journal}
  {\bibinfo  {journal} {Rev. Mod. Phys.}\ }\textbf {\bibinfo {volume} {90}},\
  \bibinfo {pages} {015001} (\bibinfo {year} {2018})}\BibitemShut {NoStop}%
\bibitem [{\citenamefont {Lv}\ \emph {et~al.}(2015{\natexlab{a}})\citenamefont
  {Lv}, \citenamefont {Xu}, \citenamefont {Weng}, \citenamefont {Ma},
  \citenamefont {Richard}, \citenamefont {Huang}, \citenamefont {Zhao},
  \citenamefont {Chen}, \citenamefont {Matt}, \citenamefont {Bisti},
  \citenamefont {Strocov}, \citenamefont {Mesot}, \citenamefont {Fang},
  \citenamefont {Dai}, \citenamefont {Qian}, \citenamefont {Shi},\ and\
  \citenamefont {Ding}}]{lv.xu.15}%
  \BibitemOpen
  \bibfield  {author} {\bibinfo {author} {\bibfnamefont {B.~Q.}\ \bibnamefont
  {Lv}}, \bibinfo {author} {\bibfnamefont {N.}~\bibnamefont {Xu}}, \bibinfo
  {author} {\bibfnamefont {H.~M.}\ \bibnamefont {Weng}}, \bibinfo {author}
  {\bibfnamefont {J.~Z.}\ \bibnamefont {Ma}}, \bibinfo {author} {\bibfnamefont
  {P.}~\bibnamefont {Richard}}, \bibinfo {author} {\bibfnamefont {X.~C.}\
  \bibnamefont {Huang}}, \bibinfo {author} {\bibfnamefont {L.~X.}\ \bibnamefont
  {Zhao}}, \bibinfo {author} {\bibfnamefont {G.~F.}\ \bibnamefont {Chen}},
  \bibinfo {author} {\bibfnamefont {C.~E.}\ \bibnamefont {Matt}}, \bibinfo
  {author} {\bibfnamefont {F.}~\bibnamefont {Bisti}}, \bibinfo {author}
  {\bibfnamefont {V.~N.}\ \bibnamefont {Strocov}}, \bibinfo {author}
  {\bibfnamefont {J.}~\bibnamefont {Mesot}}, \bibinfo {author} {\bibfnamefont
  {Z.}~\bibnamefont {Fang}}, \bibinfo {author} {\bibfnamefont {X.}~\bibnamefont
  {Dai}}, \bibinfo {author} {\bibfnamefont {T.}~\bibnamefont {Qian}}, \bibinfo
  {author} {\bibfnamefont {M.}~\bibnamefont {Shi}},\ and\ \bibinfo {author}
  {\bibfnamefont {H.}~\bibnamefont {Ding}},\ }\bibfield  {title} {\bibinfo
  {title} {Observation of {Weyl} nodes in {TaAs}},\ }\href
  {https://doi.org/10.1038/nphys3426} {\bibfield  {journal} {\bibinfo
  {journal} {Nat. Phys.}\ }\textbf {\bibinfo {volume} {11}},\ \bibinfo {pages}
  {724} (\bibinfo {year} {2015}{\natexlab{a}})}\BibitemShut {NoStop}%
\bibitem [{\citenamefont {Lv}\ \emph {et~al.}(2015{\natexlab{b}})\citenamefont
  {Lv}, \citenamefont {Weng}, \citenamefont {Fu}, \citenamefont {Wang},
  \citenamefont {Miao}, \citenamefont {Ma}, \citenamefont {Richard},
  \citenamefont {Huang}, \citenamefont {Zhao}, \citenamefont {Chen},
  \citenamefont {Fang}, \citenamefont {Dai}, \citenamefont {Qian},\ and\
  \citenamefont {Ding}}]{lv.weng.15}%
  \BibitemOpen
  \bibfield  {author} {\bibinfo {author} {\bibfnamefont {B.~Q.}\ \bibnamefont
  {Lv}}, \bibinfo {author} {\bibfnamefont {H.~M.}\ \bibnamefont {Weng}},
  \bibinfo {author} {\bibfnamefont {B.~B.}\ \bibnamefont {Fu}}, \bibinfo
  {author} {\bibfnamefont {X.~P.}\ \bibnamefont {Wang}}, \bibinfo {author}
  {\bibfnamefont {H.}~\bibnamefont {Miao}}, \bibinfo {author} {\bibfnamefont
  {J.}~\bibnamefont {Ma}}, \bibinfo {author} {\bibfnamefont {P.}~\bibnamefont
  {Richard}}, \bibinfo {author} {\bibfnamefont {X.~C.}\ \bibnamefont {Huang}},
  \bibinfo {author} {\bibfnamefont {L.~X.}\ \bibnamefont {Zhao}}, \bibinfo
  {author} {\bibfnamefont {G.~F.}\ \bibnamefont {Chen}}, \bibinfo {author}
  {\bibfnamefont {Z.}~\bibnamefont {Fang}}, \bibinfo {author} {\bibfnamefont
  {X.}~\bibnamefont {Dai}}, \bibinfo {author} {\bibfnamefont {T.}~\bibnamefont
  {Qian}},\ and\ \bibinfo {author} {\bibfnamefont {H.}~\bibnamefont {Ding}},\
  }\bibfield  {title} {\bibinfo {title} {Experimental discovery of {Weyl}
  semimetal {TaAs}},\ }\href {https://doi.org/10.1103/PhysRevX.5.031013}
  {\bibfield  {journal} {\bibinfo  {journal} {Phys. Rev. X}\ }\textbf {\bibinfo
  {volume} {5}},\ \bibinfo {pages} {031013} (\bibinfo {year}
  {2015}{\natexlab{b}})}\BibitemShut {NoStop}%
\bibitem [{\citenamefont {Xu}\ \emph {et~al.}(2015)\citenamefont {Xu},
  \citenamefont {Belopolski}, \citenamefont {Alidoust}, \citenamefont
  {Neupane}, \citenamefont {Bian}, \citenamefont {Zhang}, \citenamefont
  {Sankar}, \citenamefont {Chang}, \citenamefont {Yuan}, \citenamefont {Lee},
  \citenamefont {Huang}, \citenamefont {Zheng}, \citenamefont {Ma},
  \citenamefont {Sanchez}, \citenamefont {Wang}, \citenamefont {Bansil},
  \citenamefont {Chou}, \citenamefont {Shibayev}, \citenamefont {Lin},
  \citenamefont {Jia},\ and\ \citenamefont {Hasan}}]{xu.belopolski.15}%
  \BibitemOpen
  \bibfield  {author} {\bibinfo {author} {\bibfnamefont {S.-Y.}\ \bibnamefont
  {Xu}}, \bibinfo {author} {\bibfnamefont {I.}~\bibnamefont {Belopolski}},
  \bibinfo {author} {\bibfnamefont {N.}~\bibnamefont {Alidoust}}, \bibinfo
  {author} {\bibfnamefont {M.}~\bibnamefont {Neupane}}, \bibinfo {author}
  {\bibfnamefont {G.}~\bibnamefont {Bian}}, \bibinfo {author} {\bibfnamefont
  {C.}~\bibnamefont {Zhang}}, \bibinfo {author} {\bibfnamefont
  {R.}~\bibnamefont {Sankar}}, \bibinfo {author} {\bibfnamefont
  {G.}~\bibnamefont {Chang}}, \bibinfo {author} {\bibfnamefont
  {Z.}~\bibnamefont {Yuan}}, \bibinfo {author} {\bibfnamefont {C.-C.}\
  \bibnamefont {Lee}}, \bibinfo {author} {\bibfnamefont {S.-M.}\ \bibnamefont
  {Huang}}, \bibinfo {author} {\bibfnamefont {H.}~\bibnamefont {Zheng}},
  \bibinfo {author} {\bibfnamefont {J.}~\bibnamefont {Ma}}, \bibinfo {author}
  {\bibfnamefont {D.~S.}\ \bibnamefont {Sanchez}}, \bibinfo {author}
  {\bibfnamefont {B.}~\bibnamefont {Wang}}, \bibinfo {author} {\bibfnamefont
  {A.}~\bibnamefont {Bansil}}, \bibinfo {author} {\bibfnamefont
  {F.}~\bibnamefont {Chou}}, \bibinfo {author} {\bibfnamefont {P.~P.}\
  \bibnamefont {Shibayev}}, \bibinfo {author} {\bibfnamefont {H.}~\bibnamefont
  {Lin}}, \bibinfo {author} {\bibfnamefont {S.}~\bibnamefont {Jia}},\ and\
  \bibinfo {author} {\bibfnamefont {M.~Z.}\ \bibnamefont {Hasan}},\ }\bibfield
  {title} {\bibinfo {title} {Discovery of a {Weyl} fermion semimetal and
  topological {Fermi} arcs},\ }\href {https://doi.org/10.1126/science.aaa9297}
  {\bibfield  {journal} {\bibinfo  {journal} {Science}\ }\textbf {\bibinfo
  {volume} {349}},\ \bibinfo {pages} {613} (\bibinfo {year}
  {2015})}\BibitemShut {NoStop}%
\bibitem [{\citenamefont {Yang}\ \emph {et~al.}(2015)\citenamefont {Yang},
  \citenamefont {Liu}, \citenamefont {Sun}, \citenamefont {Peng}, \citenamefont
  {Yang}, \citenamefont {Zhang}, \citenamefont {Zhou}, \citenamefont {Zhang},
  \citenamefont {Guo}, \citenamefont {Rahn}, \citenamefont {Prabhakaran},
  \citenamefont {Hussain}, \citenamefont {Mo}, \citenamefont {Felser},
  \citenamefont {Yan},\ and\ \citenamefont {Chen}}]{yang.liu.15}%
  \BibitemOpen
  \bibfield  {author} {\bibinfo {author} {\bibfnamefont {L.~X.}\ \bibnamefont
  {Yang}}, \bibinfo {author} {\bibfnamefont {Z.~K.}\ \bibnamefont {Liu}},
  \bibinfo {author} {\bibfnamefont {Y.}~\bibnamefont {Sun}}, \bibinfo {author}
  {\bibfnamefont {H.}~\bibnamefont {Peng}}, \bibinfo {author} {\bibfnamefont
  {H.~F.}\ \bibnamefont {Yang}}, \bibinfo {author} {\bibfnamefont
  {T.}~\bibnamefont {Zhang}}, \bibinfo {author} {\bibfnamefont
  {B.}~\bibnamefont {Zhou}}, \bibinfo {author} {\bibfnamefont {Y.}~\bibnamefont
  {Zhang}}, \bibinfo {author} {\bibfnamefont {Y.~F.}\ \bibnamefont {Guo}},
  \bibinfo {author} {\bibfnamefont {M.}~\bibnamefont {Rahn}}, \bibinfo {author}
  {\bibfnamefont {D.}~\bibnamefont {Prabhakaran}}, \bibinfo {author}
  {\bibfnamefont {Z.}~\bibnamefont {Hussain}}, \bibinfo {author} {\bibfnamefont
  {S.-K.}\ \bibnamefont {Mo}}, \bibinfo {author} {\bibfnamefont
  {C.}~\bibnamefont {Felser}}, \bibinfo {author} {\bibfnamefont
  {B.}~\bibnamefont {Yan}},\ and\ \bibinfo {author} {\bibfnamefont {Y.~L.}\
  \bibnamefont {Chen}},\ }\bibfield  {title} {\bibinfo {title} {Weyl semimetal
  phase in the non-centrosymmetric compound {TaAs}},\ }\href
  {https://doi.org/10.1038/nphys3425} {\bibfield  {journal} {\bibinfo
  {journal} {Nat. Phys.}\ }\textbf {\bibinfo {volume} {11}},\ \bibinfo {pages}
  {728} (\bibinfo {year} {2015})}\BibitemShut {NoStop}%
\bibitem [{\citenamefont {Cichorek}\ \emph {et~al.}(2022)\citenamefont
  {Cichorek}, \citenamefont {Bochenek}, \citenamefont {Juraszek}, \citenamefont
  {Sharlai},\ and\ \citenamefont {Mikitik}}]{cichorek.bochenek.22}%
  \BibitemOpen
  \bibfield  {author} {\bibinfo {author} {\bibfnamefont {T.}~\bibnamefont
  {Cichorek}}, \bibinfo {author} {\bibfnamefont {L.}~\bibnamefont {Bochenek}},
  \bibinfo {author} {\bibfnamefont {J.}~\bibnamefont {Juraszek}}, \bibinfo
  {author} {\bibfnamefont {Y.~V.}\ \bibnamefont {Sharlai}},\ and\ \bibinfo
  {author} {\bibfnamefont {G.~P.}\ \bibnamefont {Mikitik}},\ }\bibfield
  {title} {\bibinfo {title} {Detection of relativistic fermions in {Weyl}
  semimetal {TaAs} by magnetostriction measurements},\ }\href
  {https://doi.org/10.1038/s41467-022-31321-4} {\bibfield  {journal} {\bibinfo
  {journal} {Nat. Commun.}\ }\textbf {\bibinfo {volume} {13}},\ \bibinfo
  {pages} {3868} (\bibinfo {year} {2022})}\BibitemShut {NoStop}%
\bibitem [{\citenamefont {Wang}\ \emph {et~al.}(2012)\citenamefont {Wang},
  \citenamefont {Sun}, \citenamefont {Chen}, \citenamefont {Franchini},
  \citenamefont {Xu}, \citenamefont {Weng}, \citenamefont {Dai},\ and\
  \citenamefont {Fang}}]{wang.sun.12}%
  \BibitemOpen
  \bibfield  {author} {\bibinfo {author} {\bibfnamefont {Z.}~\bibnamefont
  {Wang}}, \bibinfo {author} {\bibfnamefont {Y.}~\bibnamefont {Sun}}, \bibinfo
  {author} {\bibfnamefont {X.-Q.}\ \bibnamefont {Chen}}, \bibinfo {author}
  {\bibfnamefont {C.}~\bibnamefont {Franchini}}, \bibinfo {author}
  {\bibfnamefont {G.}~\bibnamefont {Xu}}, \bibinfo {author} {\bibfnamefont
  {H.}~\bibnamefont {Weng}}, \bibinfo {author} {\bibfnamefont {X.}~\bibnamefont
  {Dai}},\ and\ \bibinfo {author} {\bibfnamefont {Z.}~\bibnamefont {Fang}},\
  }\bibfield  {title} {\bibinfo {title} {Dirac semimetal and topological phase
  transitions in {${A}_{3}$Bi} ({$A=$Na, K, Rb})},\ }\href
  {https://doi.org/10.1103/PhysRevB.85.195320} {\bibfield  {journal} {\bibinfo
  {journal} {Phys. Rev. B}\ }\textbf {\bibinfo {volume} {85}},\ \bibinfo
  {pages} {195320} (\bibinfo {year} {2012})}\BibitemShut {NoStop}%
\bibitem [{\citenamefont {Liu}\ \emph {et~al.}(2014)\citenamefont {Liu},
  \citenamefont {Zhou}, \citenamefont {Zhang}, \citenamefont {Wang},
  \citenamefont {Weng}, \citenamefont {Prabhakaran}, \citenamefont {Mo},
  \citenamefont {Shen}, \citenamefont {Fang}, \citenamefont {Dai},
  \citenamefont {Hussain},\ and\ \citenamefont {Chen}}]{liu.zhou.14}%
  \BibitemOpen
  \bibfield  {author} {\bibinfo {author} {\bibfnamefont {Z.~K.}\ \bibnamefont
  {Liu}}, \bibinfo {author} {\bibfnamefont {B.}~\bibnamefont {Zhou}}, \bibinfo
  {author} {\bibfnamefont {Y.}~\bibnamefont {Zhang}}, \bibinfo {author}
  {\bibfnamefont {Z.~J.}\ \bibnamefont {Wang}}, \bibinfo {author}
  {\bibfnamefont {H.~M.}\ \bibnamefont {Weng}}, \bibinfo {author}
  {\bibfnamefont {D.}~\bibnamefont {Prabhakaran}}, \bibinfo {author}
  {\bibfnamefont {S.-K.}\ \bibnamefont {Mo}}, \bibinfo {author} {\bibfnamefont
  {Z.~X.}\ \bibnamefont {Shen}}, \bibinfo {author} {\bibfnamefont
  {Z.}~\bibnamefont {Fang}}, \bibinfo {author} {\bibfnamefont {X.}~\bibnamefont
  {Dai}}, \bibinfo {author} {\bibfnamefont {Z.}~\bibnamefont {Hussain}},\ and\
  \bibinfo {author} {\bibfnamefont {Y.~L.}\ \bibnamefont {Chen}},\ }\bibfield
  {title} {\bibinfo {title} {Discovery of a three-dimensional topological
  {Dirac} semimetal, {Na$_{3}$Bi}},\ }\href
  {https://doi.org/10.1126/science.1245085} {\bibfield  {journal} {\bibinfo
  {journal} {Science}\ }\textbf {\bibinfo {volume} {343}},\ \bibinfo {pages}
  {864} (\bibinfo {year} {2014})}\BibitemShut {NoStop}%
\bibitem [{\citenamefont {Zhang}\ \emph {et~al.}(2014)\citenamefont {Zhang},
  \citenamefont {Liu}, \citenamefont {Zhou}, \citenamefont {Kim}, \citenamefont
  {Hussain}, \citenamefont {Shen}, \citenamefont {Chen},\ and\ \citenamefont
  {Mo}}]{zhang.liu.14}%
  \BibitemOpen
  \bibfield  {author} {\bibinfo {author} {\bibfnamefont {Y.}~\bibnamefont
  {Zhang}}, \bibinfo {author} {\bibfnamefont {Z.}~\bibnamefont {Liu}}, \bibinfo
  {author} {\bibfnamefont {B.}~\bibnamefont {Zhou}}, \bibinfo {author}
  {\bibfnamefont {Y.}~\bibnamefont {Kim}}, \bibinfo {author} {\bibfnamefont
  {Z.}~\bibnamefont {Hussain}}, \bibinfo {author} {\bibfnamefont {Z.-X.}\
  \bibnamefont {Shen}}, \bibinfo {author} {\bibfnamefont {Y.}~\bibnamefont
  {Chen}},\ and\ \bibinfo {author} {\bibfnamefont {S.-K.}\ \bibnamefont {Mo}},\
  }\bibfield  {title} {\bibinfo {title} {{Molecular beam epitaxial growth of a
  three-dimensional topological Dirac semimetal {Na$_{3}$Bi}}},\ }\href
  {https://doi.org/10.1063/1.4890940} {\bibfield  {journal} {\bibinfo
  {journal} {Appl. Phys. Lett.}\ }\textbf {\bibinfo {volume} {105}},\ \bibinfo
  {pages} {031901} (\bibinfo {year} {2014})}\BibitemShut {NoStop}%
\bibitem [{\citenamefont {Wang}\ \emph {et~al.}(2013)\citenamefont {Wang},
  \citenamefont {Weng}, \citenamefont {Wu}, \citenamefont {Dai},\ and\
  \citenamefont {Fang}}]{wang.weng.13}%
  \BibitemOpen
  \bibfield  {author} {\bibinfo {author} {\bibfnamefont {Z.}~\bibnamefont
  {Wang}}, \bibinfo {author} {\bibfnamefont {H.}~\bibnamefont {Weng}}, \bibinfo
  {author} {\bibfnamefont {Q.}~\bibnamefont {Wu}}, \bibinfo {author}
  {\bibfnamefont {X.}~\bibnamefont {Dai}},\ and\ \bibinfo {author}
  {\bibfnamefont {Z.}~\bibnamefont {Fang}},\ }\bibfield  {title} {\bibinfo
  {title} {Three-dimensional {Dirac} semimetal and quantum transport in
  {Cd$_{3}$As$_{2}$}},\ }\href {https://doi.org/10.1103/PhysRevB.88.125427}
  {\bibfield  {journal} {\bibinfo  {journal} {Phys. Rev. B}\ }\textbf {\bibinfo
  {volume} {88}},\ \bibinfo {pages} {125427} (\bibinfo {year}
  {2013})}\BibitemShut {NoStop}%
\bibitem [{\citenamefont {Borisenko}\ \emph {et~al.}(2014)\citenamefont
  {Borisenko}, \citenamefont {Gibson}, \citenamefont {Evtushinsky},
  \citenamefont {Zabolotnyy}, \citenamefont {B\"uchner},\ and\ \citenamefont
  {Cava}}]{borisenko.gibson.14}%
  \BibitemOpen
  \bibfield  {author} {\bibinfo {author} {\bibfnamefont {S.}~\bibnamefont
  {Borisenko}}, \bibinfo {author} {\bibfnamefont {Q.}~\bibnamefont {Gibson}},
  \bibinfo {author} {\bibfnamefont {D.}~\bibnamefont {Evtushinsky}}, \bibinfo
  {author} {\bibfnamefont {V.}~\bibnamefont {Zabolotnyy}}, \bibinfo {author}
  {\bibfnamefont {B.}~\bibnamefont {B\"uchner}},\ and\ \bibinfo {author}
  {\bibfnamefont {R.~J.}\ \bibnamefont {Cava}},\ }\bibfield  {title} {\bibinfo
  {title} {Experimental realization of a three-dimensional {Dirac} semimetal},\
  }\href {https://doi.org/10.1103/PhysRevLett.113.027603} {\bibfield  {journal}
  {\bibinfo  {journal} {Phys. Rev. Lett.}\ }\textbf {\bibinfo {volume} {113}},\
  \bibinfo {pages} {027603} (\bibinfo {year} {2014})}\BibitemShut {NoStop}%
\bibitem [{\citenamefont {Neupane}\ \emph {et~al.}(2014)\citenamefont
  {Neupane}, \citenamefont {Xu}, \citenamefont {Sankar}, \citenamefont
  {Alidoust}, \citenamefont {Bian}, \citenamefont {Liu}, \citenamefont
  {Belopolski}, \citenamefont {Chang}, \citenamefont {Jeng}, \citenamefont
  {Lin}, \citenamefont {Bansil}, \citenamefont {Chou},\ and\ \citenamefont
  {Hasan}}]{neupane.xu.14}%
  \BibitemOpen
  \bibfield  {author} {\bibinfo {author} {\bibfnamefont {M.}~\bibnamefont
  {Neupane}}, \bibinfo {author} {\bibfnamefont {S.-Y.}\ \bibnamefont {Xu}},
  \bibinfo {author} {\bibfnamefont {R.}~\bibnamefont {Sankar}}, \bibinfo
  {author} {\bibfnamefont {N.}~\bibnamefont {Alidoust}}, \bibinfo {author}
  {\bibfnamefont {G.}~\bibnamefont {Bian}}, \bibinfo {author} {\bibfnamefont
  {C.}~\bibnamefont {Liu}}, \bibinfo {author} {\bibfnamefont {I.}~\bibnamefont
  {Belopolski}}, \bibinfo {author} {\bibfnamefont {T.-R.}\ \bibnamefont
  {Chang}}, \bibinfo {author} {\bibfnamefont {H.-T.}\ \bibnamefont {Jeng}},
  \bibinfo {author} {\bibfnamefont {H.}~\bibnamefont {Lin}}, \bibinfo {author}
  {\bibfnamefont {A.}~\bibnamefont {Bansil}}, \bibinfo {author} {\bibfnamefont
  {F.}~\bibnamefont {Chou}},\ and\ \bibinfo {author} {\bibfnamefont {M.~Z.}\
  \bibnamefont {Hasan}},\ }\bibfield  {title} {\bibinfo {title} {Observation of
  a three-dimensional topological {Dirac} semimetal phase in high-mobility
  {Cd$_{3}$As$_{2}$}},\ }\href {https://doi.org/10.1038/ncomms4786} {\bibfield
  {journal} {\bibinfo  {journal} {Nat. Commun.}\ }\textbf {\bibinfo {volume}
  {5}},\ \bibinfo {pages} {3786} (\bibinfo {year} {2014})}\BibitemShut
  {NoStop}%
\bibitem [{\citenamefont {Ruan}\ \emph {et~al.}(2016)\citenamefont {Ruan},
  \citenamefont {Jian}, \citenamefont {Yao}, \citenamefont {Zhang},
  \citenamefont {Zhang},\ and\ \citenamefont {Xing}}]{ruan.jian.16}%
  \BibitemOpen
  \bibfield  {author} {\bibinfo {author} {\bibfnamefont {J.}~\bibnamefont
  {Ruan}}, \bibinfo {author} {\bibfnamefont {S.-K.}\ \bibnamefont {Jian}},
  \bibinfo {author} {\bibfnamefont {H.}~\bibnamefont {Yao}}, \bibinfo {author}
  {\bibfnamefont {H.}~\bibnamefont {Zhang}}, \bibinfo {author} {\bibfnamefont
  {S.-C.}\ \bibnamefont {Zhang}},\ and\ \bibinfo {author} {\bibfnamefont
  {D.}~\bibnamefont {Xing}},\ }\bibfield  {title} {\bibinfo {title}
  {Symmetry-protected ideal {Weyl} semimetal in {HgTe}-class materials},\
  }\href {https://doi.org/10.1038/ncomms11136} {\bibfield  {journal} {\bibinfo
  {journal} {Nat. Commun.}\ }\textbf {\bibinfo {volume} {7}},\ \bibinfo {pages}
  {11136} (\bibinfo {year} {2016})}\BibitemShut {NoStop}%
\bibitem [{\citenamefont {Ferreira}\ \emph {et~al.}(2021)\citenamefont
  {Ferreira}, \citenamefont {Manesco}, \citenamefont {Dorini}, \citenamefont
  {Correa}, \citenamefont {Weber}, \citenamefont {Machado},\ and\ \citenamefont
  {Eleno}}]{ferreira.manesco.21}%
  \BibitemOpen
  \bibfield  {author} {\bibinfo {author} {\bibfnamefont {P.~P.}\ \bibnamefont
  {Ferreira}}, \bibinfo {author} {\bibfnamefont {A.~L.~R.}\ \bibnamefont
  {Manesco}}, \bibinfo {author} {\bibfnamefont {T.~T.}\ \bibnamefont {Dorini}},
  \bibinfo {author} {\bibfnamefont {L.~E.}\ \bibnamefont {Correa}}, \bibinfo
  {author} {\bibfnamefont {G.}~\bibnamefont {Weber}}, \bibinfo {author}
  {\bibfnamefont {A.~J.~S.}\ \bibnamefont {Machado}},\ and\ \bibinfo {author}
  {\bibfnamefont {L.~T.~F.}\ \bibnamefont {Eleno}},\ }\bibfield  {title}
  {\bibinfo {title} {Strain engineering the topological type-{II} {Dirac}
  semimetal {NiTe$_{2}$}},\ }\href
  {https://doi.org/10.1103/PhysRevB.103.125134} {\bibfield  {journal} {\bibinfo
   {journal} {Phys. Rev. B}\ }\textbf {\bibinfo {volume} {103}},\ \bibinfo
  {pages} {125134} (\bibinfo {year} {2021})}\BibitemShut {NoStop}%
\bibitem [{\citenamefont {Wu}\ \emph {et~al.}(2021)\citenamefont {Wu},
  \citenamefont {Li}, \citenamefont {Zhao}, \citenamefont {Dai}, \citenamefont
  {Zhao},\ and\ \citenamefont {Meng}}]{wu.li.21}%
  \BibitemOpen
  \bibfield  {author} {\bibinfo {author} {\bibfnamefont {J.}~\bibnamefont
  {Wu}}, \bibinfo {author} {\bibfnamefont {Y.}~\bibnamefont {Li}}, \bibinfo
  {author} {\bibfnamefont {L.}~\bibnamefont {Zhao}}, \bibinfo {author}
  {\bibfnamefont {T.}~\bibnamefont {Dai}}, \bibinfo {author} {\bibfnamefont
  {X.}~\bibnamefont {Zhao}},\ and\ \bibinfo {author} {\bibfnamefont
  {L.}~\bibnamefont {Meng}},\ }\bibfield  {title} {\bibinfo {title} {The
  monolayer alloying and strain effect in {Weyl} semimetal {Td-MoTe$_{2}$}},\
  }\href {https://doi.org/10.1016/j.jpcs.2020.109739} {\bibfield  {journal}
  {\bibinfo  {journal} {J. Phys. Chem. Solids}\ }\textbf {\bibinfo {volume}
  {148}},\ \bibinfo {pages} {109739} (\bibinfo {year} {2021})}\BibitemShut
  {NoStop}%
\bibitem [{\citenamefont {Li}\ \emph {et~al.}(2022)\citenamefont {Li},
  \citenamefont {Zhao}, \citenamefont {Zhao}, \citenamefont {Dai},
  \citenamefont {Zhong},\ and\ \citenamefont {Meng}}]{li.zhao.22}%
  \BibitemOpen
  \bibfield  {author} {\bibinfo {author} {\bibfnamefont {Y.}~\bibnamefont
  {Li}}, \bibinfo {author} {\bibfnamefont {L.}~\bibnamefont {Zhao}}, \bibinfo
  {author} {\bibfnamefont {X.}~\bibnamefont {Zhao}}, \bibinfo {author}
  {\bibfnamefont {T.}~\bibnamefont {Dai}}, \bibinfo {author} {\bibfnamefont
  {J.}~\bibnamefont {Zhong}},\ and\ \bibinfo {author} {\bibfnamefont
  {L.}~\bibnamefont {Meng}},\ }\bibfield  {title} {\bibinfo {title} {Magnetic
  field and strain effects in {Janus}-like {Weyl} semimetals {MoTeSe} with four
  {Weyl} points},\ }\href {https://doi.org/10.1016/j.commatsci.2022.111617}
  {\bibfield  {journal} {\bibinfo  {journal} {Comput. Mater. Sci.}\ }\textbf
  {\bibinfo {volume} {213}},\ \bibinfo {pages} {111617} (\bibinfo {year}
  {2022})}\BibitemShut {NoStop}%
\bibitem [{\citenamefont {Wu}\ \emph {et~al.}(2023)\citenamefont {Wu},
  \citenamefont {Ke}, \citenamefont {Guo}, \citenamefont {Zhang},\ and\
  \citenamefont {Lü}}]{wu.ke.23}%
  \BibitemOpen
  \bibfield  {author} {\bibinfo {author} {\bibfnamefont {J.-F.}\ \bibnamefont
  {Wu}}, \bibinfo {author} {\bibfnamefont {S.-S.}\ \bibnamefont {Ke}}, \bibinfo
  {author} {\bibfnamefont {Y.}~\bibnamefont {Guo}}, \bibinfo {author}
  {\bibfnamefont {H.-W.}\ \bibnamefont {Zhang}},\ and\ \bibinfo {author}
  {\bibfnamefont {H.-F.}\ \bibnamefont {Lü}},\ }\bibfield  {title} {\bibinfo
  {title} {Non-centrosymmetric {Weyl} semimetal state and strain effect in the
  twisted-brick phase transition metal monochalcogenides},\ }\href
  {https://doi.org/10.1039/D2NR04946E} {\bibfield  {journal} {\bibinfo
  {journal} {Nanoscale}\ }\textbf {\bibinfo {volume} {15}},\ \bibinfo {pages}
  {2882} (\bibinfo {year} {2023})}\BibitemShut {NoStop}%
\bibitem [{\citenamefont {Mutch}\ \emph {et~al.}(2019)\citenamefont {Mutch},
  \citenamefont {Chen}, \citenamefont {Went}, \citenamefont {Qian},
  \citenamefont {Wilson}, \citenamefont {Andreev}, \citenamefont {Chen},\ and\
  \citenamefont {Chu}}]{mutch.chen.19}%
  \BibitemOpen
  \bibfield  {author} {\bibinfo {author} {\bibfnamefont {J.}~\bibnamefont
  {Mutch}}, \bibinfo {author} {\bibfnamefont {W.-C.}\ \bibnamefont {Chen}},
  \bibinfo {author} {\bibfnamefont {P.}~\bibnamefont {Went}}, \bibinfo {author}
  {\bibfnamefont {T.}~\bibnamefont {Qian}}, \bibinfo {author} {\bibfnamefont
  {I.~Z.}\ \bibnamefont {Wilson}}, \bibinfo {author} {\bibfnamefont
  {A.}~\bibnamefont {Andreev}}, \bibinfo {author} {\bibfnamefont {C.-C.}\
  \bibnamefont {Chen}},\ and\ \bibinfo {author} {\bibfnamefont {J.-H.}\
  \bibnamefont {Chu}},\ }\bibfield  {title} {\bibinfo {title} {Evidence for a
  strain-tuned topological phase transition in {ZrTe$_{5}$}},\ }\href
  {https://doi.org/10.1126/sciadv.aav9771} {\bibfield  {journal} {\bibinfo
  {journal} {Sci. Adv.}\ }\textbf {\bibinfo {volume} {5}},\ \bibinfo {pages}
  {eaav9771} (\bibinfo {year} {2019})}\BibitemShut {NoStop}%
\bibitem [{\citenamefont {Jiang}\ \emph {et~al.}(2021)\citenamefont {Jiang},
  \citenamefont {Guo}, \citenamefont {Wang}, \citenamefont {Wan},\ and\
  \citenamefont {Li}}]{jiang.guo.21}%
  \BibitemOpen
  \bibfield  {author} {\bibinfo {author} {\bibfnamefont {W.}~\bibnamefont
  {Jiang}}, \bibinfo {author} {\bibfnamefont {Y.}~\bibnamefont {Guo}}, \bibinfo
  {author} {\bibfnamefont {X.}~\bibnamefont {Wang}}, \bibinfo {author}
  {\bibfnamefont {F.}~\bibnamefont {Wan}},\ and\ \bibinfo {author}
  {\bibfnamefont {Y.}~\bibnamefont {Li}},\ }\bibfield  {title} {\bibinfo
  {title} {Strain modulation of the transport properties of {Weyl} semimetal
  {TaAs}},\ }\href {https://doi.org/10.1016/j.physe.2020.114600} {\bibfield
  {journal} {\bibinfo  {journal} {Physica E}\ }\textbf {\bibinfo {volume}
  {128}},\ \bibinfo {pages} {114600} (\bibinfo {year} {2021})}\BibitemShut
  {NoStop}%
\bibitem [{\citenamefont {Yip}\ \emph {et~al.}(2023)\citenamefont {Yip},
  \citenamefont {Lam}, \citenamefont {Yu}, \citenamefont {Chow}, \citenamefont
  {Zeng}, \citenamefont {Lai},\ and\ \citenamefont {Goh}}]{yip.lam.23}%
  \BibitemOpen
  \bibfield  {author} {\bibinfo {author} {\bibfnamefont {K.~Y.}\ \bibnamefont
  {Yip}}, \bibinfo {author} {\bibfnamefont {S.~T.}\ \bibnamefont {Lam}},
  \bibinfo {author} {\bibfnamefont {K.~H.}\ \bibnamefont {Yu}}, \bibinfo
  {author} {\bibfnamefont {W.~S.}\ \bibnamefont {Chow}}, \bibinfo {author}
  {\bibfnamefont {J.}~\bibnamefont {Zeng}}, \bibinfo {author} {\bibfnamefont
  {K.~T.}\ \bibnamefont {Lai}},\ and\ \bibinfo {author} {\bibfnamefont {S.~K.}\
  \bibnamefont {Goh}},\ }\bibfield  {title} {\bibinfo {title} {{Drastic
  enhancement of the superconducting temperature in type-{II} {Weyl} semimetal
  candidate {MoTe$_{2}$} via biaxial strain}},\ }\href
  {https://doi.org/10.1063/5.0141112} {\bibfield  {journal} {\bibinfo
  {journal} {APL Materials}\ }\textbf {\bibinfo {volume} {11}},\ \bibinfo
  {pages} {021111} (\bibinfo {year} {2023})}\BibitemShut {NoStop}%
\bibitem [{\citenamefont {Grushin}\ \emph {et~al.}(2016)\citenamefont
  {Grushin}, \citenamefont {Venderbos}, \citenamefont {Vishwanath},\ and\
  \citenamefont {Ilan}}]{grushin.venderbos.16}%
  \BibitemOpen
  \bibfield  {author} {\bibinfo {author} {\bibfnamefont {A.~G.}\ \bibnamefont
  {Grushin}}, \bibinfo {author} {\bibfnamefont {J.~W.~F.}\ \bibnamefont
  {Venderbos}}, \bibinfo {author} {\bibfnamefont {A.}~\bibnamefont
  {Vishwanath}},\ and\ \bibinfo {author} {\bibfnamefont {R.}~\bibnamefont
  {Ilan}},\ }\bibfield  {title} {\bibinfo {title} {Inhomogeneous weyl and dirac
  semimetals: Transport in axial magnetic fields and fermi arc surface states
  from pseudo-landau levels},\ }\href
  {https://doi.org/10.1103/PhysRevX.6.041046} {\bibfield  {journal} {\bibinfo
  {journal} {Phys. Rev. X}\ }\textbf {\bibinfo {volume} {6}},\ \bibinfo {pages}
  {041046} (\bibinfo {year} {2016})}\BibitemShut {NoStop}%
\bibitem [{\citenamefont {Liu}\ \emph {et~al.}(2013)\citenamefont {Liu},
  \citenamefont {Ye},\ and\ \citenamefont {Qi}}]{liu.ye.13}%
  \BibitemOpen
  \bibfield  {author} {\bibinfo {author} {\bibfnamefont {C.-X.}\ \bibnamefont
  {Liu}}, \bibinfo {author} {\bibfnamefont {P.}~\bibnamefont {Ye}},\ and\
  \bibinfo {author} {\bibfnamefont {X.-L.}\ \bibnamefont {Qi}},\ }\bibfield
  {title} {\bibinfo {title} {Chiral gauge field and axial anomaly in a weyl
  semimetal},\ }\href {https://doi.org/10.1103/PhysRevB.87.235306} {\bibfield
  {journal} {\bibinfo  {journal} {Phys. Rev. B}\ }\textbf {\bibinfo {volume}
  {87}},\ \bibinfo {pages} {235306} (\bibinfo {year} {2013})}\BibitemShut
  {NoStop}%
\bibitem [{\citenamefont {Sumiyoshi}\ and\ \citenamefont
  {Fujimoto}(2016)}]{sumiyoshi.fujimoto.16}%
  \BibitemOpen
  \bibfield  {author} {\bibinfo {author} {\bibfnamefont {H.}~\bibnamefont
  {Sumiyoshi}}\ and\ \bibinfo {author} {\bibfnamefont {S.}~\bibnamefont
  {Fujimoto}},\ }\bibfield  {title} {\bibinfo {title} {Torsional chiral
  magnetic effect in a {Weyl} semimetal with a topological defect},\ }\href
  {https://doi.org/10.1103/PhysRevLett.116.166601} {\bibfield  {journal}
  {\bibinfo  {journal} {Phys. Rev. Lett.}\ }\textbf {\bibinfo {volume} {116}},\
  \bibinfo {pages} {166601} (\bibinfo {year} {2016})}\BibitemShut {NoStop}%
\bibitem [{\citenamefont {Abdulla}(2023)}]{faruk.23}%
  \BibitemOpen
  \bibfield  {author} {\bibinfo {author} {\bibfnamefont {F.}~\bibnamefont
  {Abdulla}},\ }\href@noop {} {\bibinfo {title} {Pairwise annihilation of
  {Weyl} nodes induced by magnetic fields in the hofstadter regime}} (\bibinfo
  {year} {2023}),\ \Eprint {https://arxiv.org/abs/arXiv:2312.02463}
  {arXiv:2312.02463} \BibitemShut {NoStop}%
\bibitem [{\citenamefont {Chan}\ and\ \citenamefont {Lee}(2017)}]{chan.lee.17}%
  \BibitemOpen
  \bibfield  {author} {\bibinfo {author} {\bibfnamefont {C.-K.}\ \bibnamefont
  {Chan}}\ and\ \bibinfo {author} {\bibfnamefont {P.~A.}\ \bibnamefont {Lee}},\
  }\bibfield  {title} {\bibinfo {title} {Emergence of gapped bulk and metallic
  side walls in the zeroth landau level in dirac and weyl semimetals},\ }\href
  {https://doi.org/10.1103/PhysRevB.96.195143} {\bibfield  {journal} {\bibinfo
  {journal} {Phys. Rev. B}\ }\textbf {\bibinfo {volume} {96}},\ \bibinfo
  {pages} {195143} (\bibinfo {year} {2017})}\BibitemShut {NoStop}%
\bibitem [{\citenamefont {Li}\ and\ \citenamefont
  {Carbotte}(2013)}]{li.carbotte.13}%
  \BibitemOpen
  \bibfield  {author} {\bibinfo {author} {\bibfnamefont {Z.}~\bibnamefont
  {Li}}\ and\ \bibinfo {author} {\bibfnamefont {J.~P.}\ \bibnamefont
  {Carbotte}},\ }\bibfield  {title} {\bibinfo {title} {Magneto-optical
  conductivity in a topological insulator},\ }\href
  {https://doi.org/10.1103/PhysRevB.88.045414} {\bibfield  {journal} {\bibinfo
  {journal} {Phys. Rev. B}\ }\textbf {\bibinfo {volume} {88}},\ \bibinfo
  {pages} {045414} (\bibinfo {year} {2013})}\BibitemShut {NoStop}%
\bibitem [{\citenamefont {Zhao}\ \emph {et~al.}(2022)\citenamefont {Zhao},
  \citenamefont {Sun}, \citenamefont {Wang},\ and\ \citenamefont
  {Pan}}]{zhao.sun.22}%
  \BibitemOpen
  \bibfield  {author} {\bibinfo {author} {\bibfnamefont {H.}~\bibnamefont
  {Zhao}}, \bibinfo {author} {\bibfnamefont {Y.}~\bibnamefont {Sun}}, \bibinfo
  {author} {\bibfnamefont {H.}~\bibnamefont {Wang}},\ and\ \bibinfo {author}
  {\bibfnamefont {H.}~\bibnamefont {Pan}},\ }\bibfield  {title} {\bibinfo
  {title} {{Magneto-optical conductivity of nodal link semimetals}},\ }\href
  {https://doi.org/10.1063/5.0125664} {\bibfield  {journal} {\bibinfo
  {journal} {J. Appl. Phys.}\ }\textbf {\bibinfo {volume} {132}},\ \bibinfo
  {pages} {193903} (\bibinfo {year} {2022})}\BibitemShut {NoStop}%
\bibitem [{\citenamefont {Roy}\ \emph {et~al.}(2017)\citenamefont {Roy},
  \citenamefont {Goswami},\ and\ \citenamefont
  {Juri\v{c}i\'{c}}}]{roy.goswami.17}%
  \BibitemOpen
  \bibfield  {author} {\bibinfo {author} {\bibfnamefont {B.}~\bibnamefont
  {Roy}}, \bibinfo {author} {\bibfnamefont {P.}~\bibnamefont {Goswami}},\ and\
  \bibinfo {author} {\bibfnamefont {V.}~\bibnamefont {Juri\v{c}i\'{c}}},\
  }\bibfield  {title} {\bibinfo {title} {Interacting {Weyl} fermions: Phases,
  phase transitions, and global phase diagram},\ }\href
  {https://doi.org/10.1103/PhysRevB.95.201102} {\bibfield  {journal} {\bibinfo
  {journal} {Phys. Rev. B}\ }\textbf {\bibinfo {volume} {95}},\ \bibinfo
  {pages} {201102(R)} (\bibinfo {year} {2017})}\BibitemShut {NoStop}%
\bibitem [{\citenamefont {Nag}\ \emph {et~al.}(2020)\citenamefont {Nag},
  \citenamefont {Menon},\ and\ \citenamefont {Basu}}]{nag.menon.20}%
  \BibitemOpen
  \bibfield  {author} {\bibinfo {author} {\bibfnamefont {T.}~\bibnamefont
  {Nag}}, \bibinfo {author} {\bibfnamefont {A.}~\bibnamefont {Menon}},\ and\
  \bibinfo {author} {\bibfnamefont {B.}~\bibnamefont {Basu}},\ }\bibfield
  {title} {\bibinfo {title} {Thermoelectric transport properties of {Floquet}
  multi-{Weyl} semimetals},\ }\href
  {https://doi.org/10.1103/PhysRevB.102.014307} {\bibfield  {journal} {\bibinfo
   {journal} {Phys. Rev. B}\ }\textbf {\bibinfo {volume} {102}},\ \bibinfo
  {pages} {014307} (\bibinfo {year} {2020})}\BibitemShut {NoStop}%
\bibitem [{Note1()}]{Note1}%
  \BibitemOpen
  \bibinfo {note} {Details of the model can be found in Supplemental Material
  of Ref.~\cite {nag.menon.20}.}\BibitemShut {Stop}%
\bibitem [{\citenamefont {Nag}\ and\ \citenamefont
  {Nandy}(2020)}]{nag.nandy.20}%
  \BibitemOpen
  \bibfield  {author} {\bibinfo {author} {\bibfnamefont {T.}~\bibnamefont
  {Nag}}\ and\ \bibinfo {author} {\bibfnamefont {S.}~\bibnamefont {Nandy}},\
  }\bibfield  {title} {\bibinfo {title} {Magneto-transport phenomena of
  type-{I} multi-{Weyl} semimetals in co-planar setups},\ }\href
  {https://doi.org/10.1088/1361-648X/abc310} {\bibfield  {journal} {\bibinfo
  {journal} {J. Phys.: Condens. Matter}\ }\textbf {\bibinfo {volume} {33}},\
  \bibinfo {pages} {075504} (\bibinfo {year} {2020})}\BibitemShut {NoStop}%
\bibitem [{\citenamefont {Peierls}(1933)}]{peierls.33}%
  \BibitemOpen
  \bibfield  {author} {\bibinfo {author} {\bibfnamefont {R.}~\bibnamefont
  {Peierls}},\ }\bibfield  {title} {\bibinfo {title} {Z. physik},\ }\href
  {https://doi.org/10.1007/BF01342591} {\bibfield  {journal} {\bibinfo
  {journal} {Z. Physik}\ }\textbf {\bibinfo {volume} {80}},\ \bibinfo {pages}
  {763} (\bibinfo {year} {1933})}\BibitemShut {NoStop}%
\bibitem [{\citenamefont {Harper}(1955)}]{harper.55}%
  \BibitemOpen
  \bibfield  {author} {\bibinfo {author} {\bibfnamefont {P.~G.}\ \bibnamefont
  {Harper}},\ }\bibfield  {title} {\bibinfo {title} {The general motion of
  conduction electrons in a uniform magnetic field, with application to the
  diamagnetism of metals},\ }\href
  {https://doi.org/10.1088/0370-1298/68/10/305} {\bibfield  {journal} {\bibinfo
   {journal} {Proc. Phys. Soc. A}\ }\textbf {\bibinfo {volume} {68}},\ \bibinfo
  {pages} {879} (\bibinfo {year} {1955})}\BibitemShut {NoStop}%
\bibitem [{\citenamefont {Hofstadter}(1976)}]{hofstadter.76}%
  \BibitemOpen
  \bibfield  {author} {\bibinfo {author} {\bibfnamefont {D.~R.}\ \bibnamefont
  {Hofstadter}},\ }\bibfield  {title} {\bibinfo {title} {Energy levels and wave
  functions of {Bloch} electrons in rational and irrational magnetic fields},\
  }\href {https://doi.org/10.1103/PhysRevB.14.2239} {\bibfield  {journal}
  {\bibinfo  {journal} {Phys. Rev. B}\ }\textbf {\bibinfo {volume} {14}},\
  \bibinfo {pages} {2239} (\bibinfo {year} {1976})}\BibitemShut {NoStop}%
\bibitem [{Note2()}]{Note2}%
  \BibitemOpen
  \bibinfo {note} {See Supplemental Material at [URL will be inserted by
  publisher] for additional theoretical results.}\BibitemShut {Stop}%
\bibitem [{\citenamefont {Huang}\ \emph {et~al.}(2018)\citenamefont {Huang},
  \citenamefont {Jin},\ and\ \citenamefont {Liu}}]{huang.jin.18}%
  \BibitemOpen
  \bibfield  {author} {\bibinfo {author} {\bibfnamefont {H.}~\bibnamefont
  {Huang}}, \bibinfo {author} {\bibfnamefont {K.-H.}\ \bibnamefont {Jin}},\
  and\ \bibinfo {author} {\bibfnamefont {F.}~\bibnamefont {Liu}},\ }\bibfield
  {title} {\bibinfo {title} {Black-hole horizon in the {Dirac} semimetal
  {Zn$_{2}$In$_{2}$S$_{5}$}},\ }\href
  {https://doi.org/10.1103/PhysRevB.98.121110} {\bibfield  {journal} {\bibinfo
  {journal} {Phys. Rev. B}\ }\textbf {\bibinfo {volume} {98}},\ \bibinfo
  {pages} {121110(R)} (\bibinfo {year} {2018})}\BibitemShut {NoStop}%
\bibitem [{\citenamefont {Liu}\ \emph {et~al.}(2018)\citenamefont {Liu},
  \citenamefont {Sun}, \citenamefont {Cheng}, \citenamefont {Liu},\ and\
  \citenamefont {Meng}}]{liu.sun.18}%
  \BibitemOpen
  \bibfield  {author} {\bibinfo {author} {\bibfnamefont {H.}~\bibnamefont
  {Liu}}, \bibinfo {author} {\bibfnamefont {J.-T.}\ \bibnamefont {Sun}},
  \bibinfo {author} {\bibfnamefont {C.}~\bibnamefont {Cheng}}, \bibinfo
  {author} {\bibfnamefont {F.}~\bibnamefont {Liu}},\ and\ \bibinfo {author}
  {\bibfnamefont {S.}~\bibnamefont {Meng}},\ }\bibfield  {title} {\bibinfo
  {title} {Photoinduced nonequilibrium topological states in strained black
  phosphorus},\ }\href {https://doi.org/10.1103/PhysRevLett.120.237403}
  {\bibfield  {journal} {\bibinfo  {journal} {Phys. Rev. Lett.}\ }\textbf
  {\bibinfo {volume} {120}},\ \bibinfo {pages} {237403} (\bibinfo {year}
  {2018})}\BibitemShut {NoStop}%
\bibitem [{\citenamefont {Fragkos}\ \emph {et~al.}(2021)\citenamefont
  {Fragkos}, \citenamefont {Tsipas}, \citenamefont {Xenogiannopoulou},
  \citenamefont {Panayiotatos},\ and\ \citenamefont
  {Dimoulas}}]{fragkos.tsipas.21}%
  \BibitemOpen
  \bibfield  {author} {\bibinfo {author} {\bibfnamefont {S.}~\bibnamefont
  {Fragkos}}, \bibinfo {author} {\bibfnamefont {P.}~\bibnamefont {Tsipas}},
  \bibinfo {author} {\bibfnamefont {E.}~\bibnamefont {Xenogiannopoulou}},
  \bibinfo {author} {\bibfnamefont {Y.}~\bibnamefont {Panayiotatos}},\ and\
  \bibinfo {author} {\bibfnamefont {A.}~\bibnamefont {Dimoulas}},\ }\bibfield
  {title} {\bibinfo {title} {{Type-{III} {Dirac} fermions in
  {Hf$_{x}$Zr$_{1-x}$Te$_{2}$} topological semimetal candidate}},\ }\href
  {https://doi.org/10.1063/5.0038799} {\bibfield  {journal} {\bibinfo
  {journal} {J. Appl. Phys.}\ }\textbf {\bibinfo {volume} {129}},\ \bibinfo
  {pages} {075104} (\bibinfo {year} {2021})}\BibitemShut {NoStop}%
\end{thebibliography}%


	\clearpage
	\newpage

	\onecolumngrid

	\begin{center}
		\textbf{\Large Supplemental Material}\\[.3cm]
		\textbf{\large Landau Levels in Weyl semimetal under uniaxial strain}\\[.3cm]
		Shivam Yadav and Andrzej Ptok\\[.2cm]
		{\itshape
			Institute of Nuclear Physics, Polish Academy of Sciences, W. E. Radzikowskiego 152, PL-31342 Krak\'{o}w, Poland
		}
		(Dated: \today)
		\\[1cm]
	\end{center}

	\setcounter{equation}{0}
	\renewcommand{\theequation}{SE\arabic{equation}}
	\setcounter{figure}{0}
	\renewcommand{\thefigure}{SF\arabic{figure}}
	\setcounter{section}{0}
	\renewcommand{\thesection}{SS\arabic{section}}
	\setcounter{table}{0}
	\renewcommand{\thetable}{ST\arabic{table}}
	\setcounter{page}{1}

\begin{figure}[!h]
		\centering
		\includegraphics[width=0.75\linewidth]{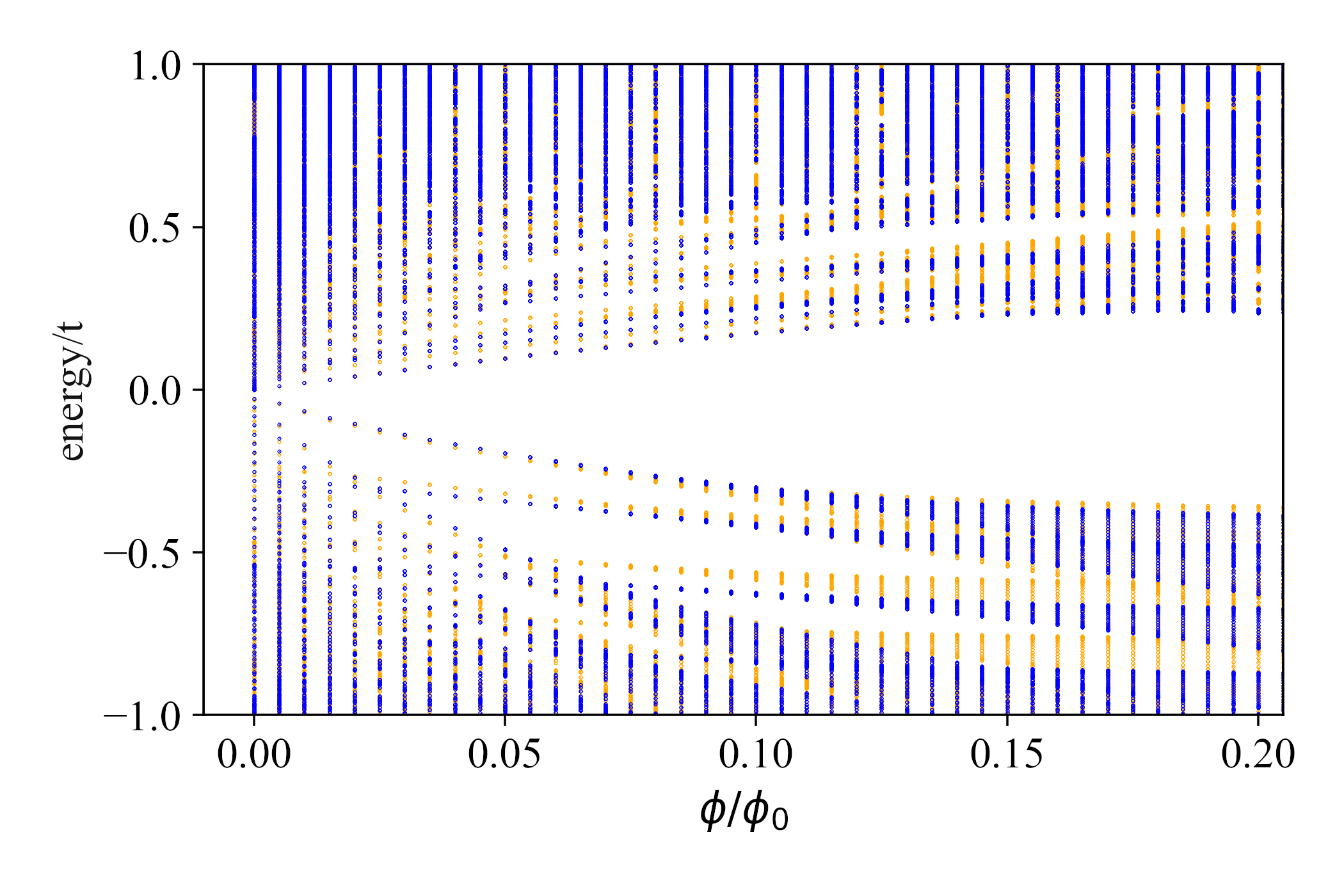}
		\caption{
			Energy spectrum of type-I model as the function of magnetic flux $\phi$ in the absence ($\alpha = 0$) and presence ($\alpha = 0.15$) uniaxial strain (orange and blue dots, respectively).
			Results for type-I WSM  for $k_z = \pm \pi / 2$ closest to the Weyl node, with $t_z = t$ and $t_0 = 0.4 t$ in the system $200 \times 200 \times 13$ sites.
			\label{fig.strain2}
		}
	\end{figure}

\begin{figure}[!t]
		\centering
		\includegraphics[width=\linewidth]{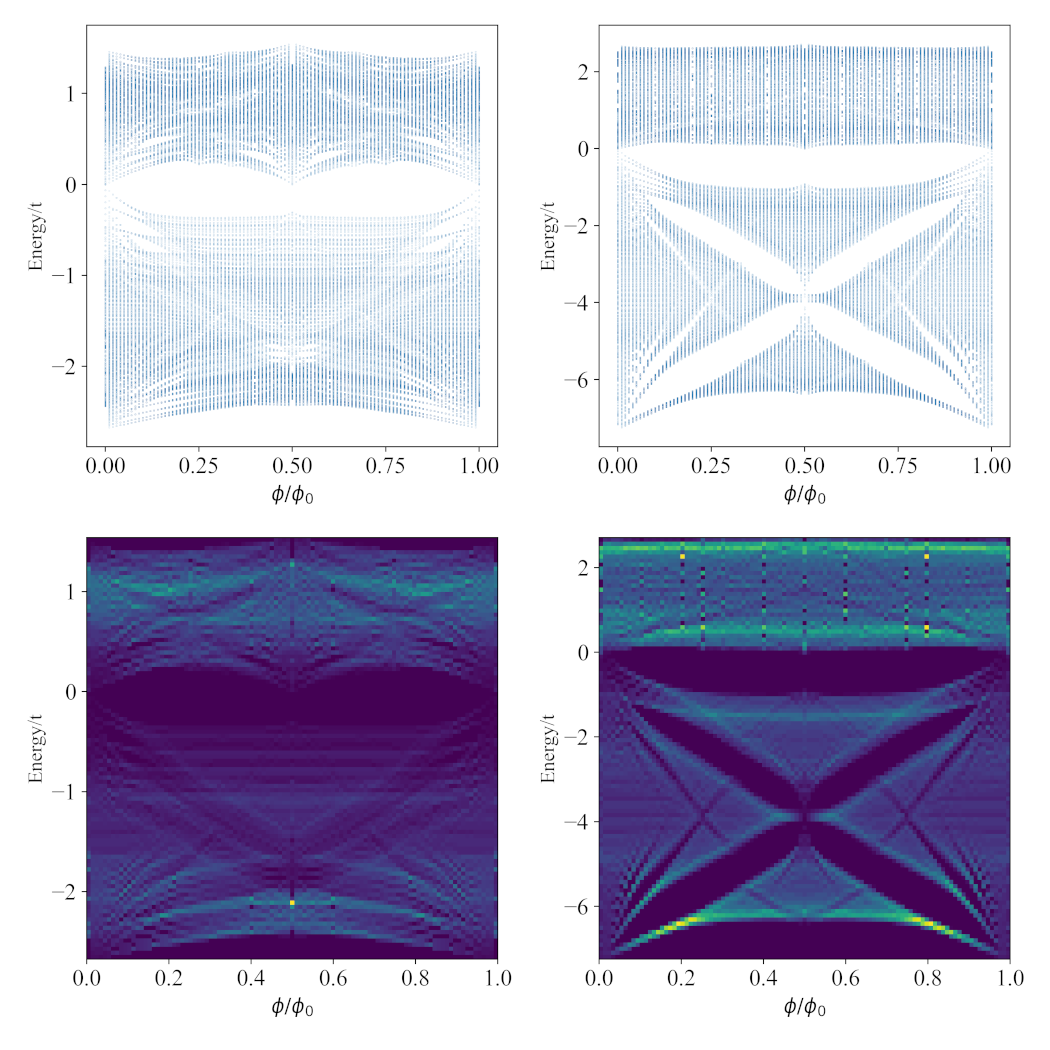}
		\caption{
			The Hofstadter butterfly (top row) and the density of states (bottom row) for type-I and type-II (left and right column, respectively).
			Results for the slices nearest to $k_z = \pm \pi/2$ without uniaxial strain, and lattice size $(100 \times 100 \times 13)$.
			For the type-I we set $t_0 = 0.4 t$, while for type-II $t_{0} = 1.2 t$.
			\label{fig.butterflies}
		}
	\end{figure}

\begin{figure}[!t]
		\centering
		\includegraphics[width=\linewidth]{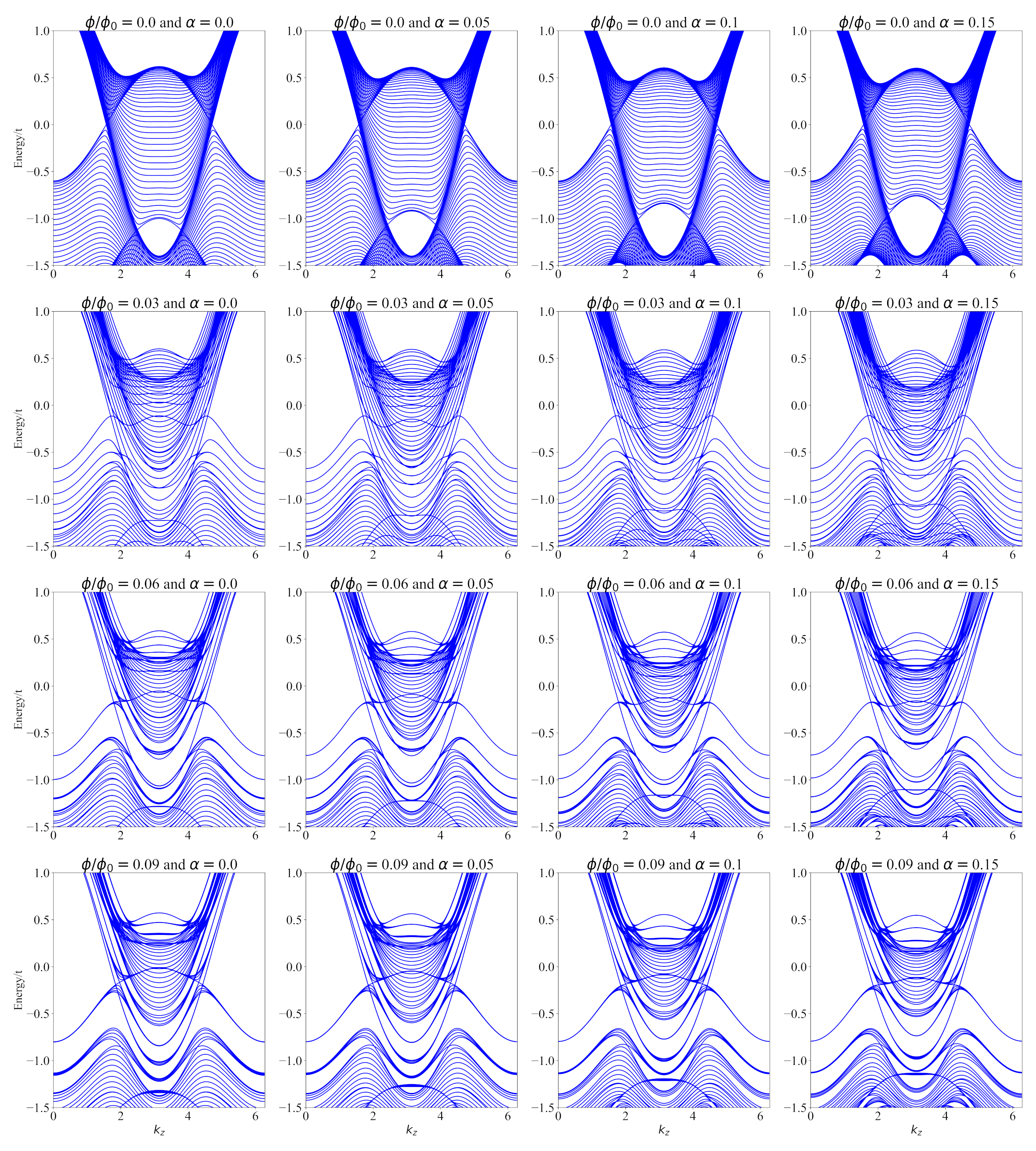}
		\caption{
			Comparison of the band structure for the type-I lattice model under uniaxial strain $\alpha$ (columns from left to right) and magnetic field $\phi$ (rows from top to bottom).
			Results for momenta $k_y = 0$ which is the closest to the Weyl node.
			\label{fig.disp_combined}
		}
	\end{figure}

\begin{figure*}
		\centering
		\includegraphics[width=\linewidth]{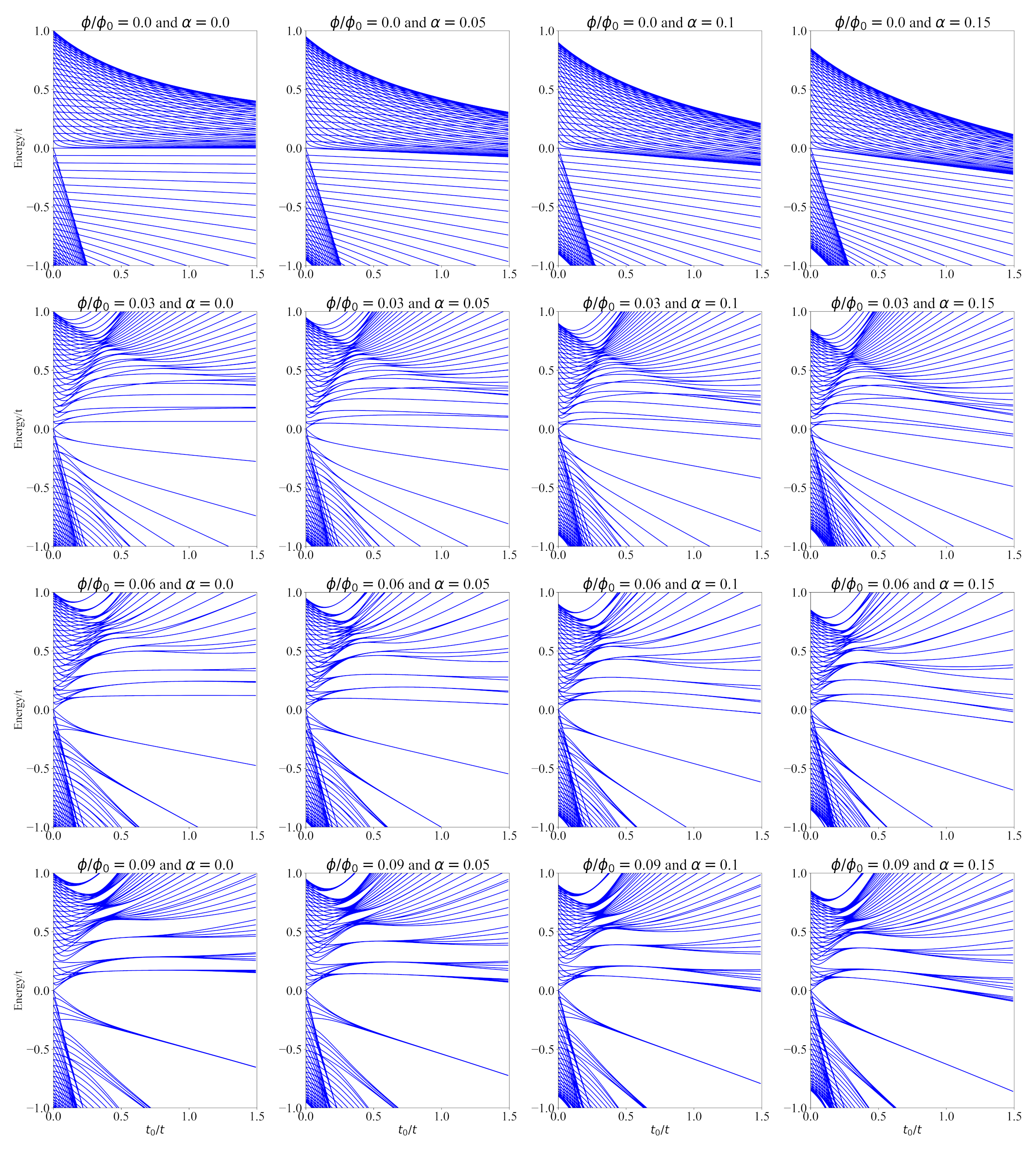}
		\caption{
			Panel illustrates the energy spectrum as the function of tilt (controled by $t_{0}$ parameter) for fixed magnetic flux $\phi$ and uniaxial strain $\alpha$ (as labelled).
			For the critical $t_{0} = t_{z}$ (in our case $t_{0} = t$), the Weyl node changes from type-I to type-II.
			Results for $k_y = 0$ and $k_z = \pi/2$, which are the closest to the Weyl node.
			\label{fig.disp_to}
		}
	\end{figure*}

	In this Supplemental Material, we present additional results:
	\begin{itemize}
		\item Fig.~\ref{fig.strain2} -- Comparison of the energy spectrum in the absence and presence of uniaxial strain.
		\item Fig.~\ref{fig.butterflies} -- The Hofstadter butterflies for type-I and type-II Weyl semimetal model.
		\item Fig.~\ref{fig.disp_combined} -- The band structures in the presence of magnetic flux and uniaxial strain.
		\item Fig.~\ref{fig.disp_to} -- System spectrum with respect to strain and magnetic flux.
	\end{itemize}

\end{document}